\documentclass[aps,prd,floatfix,superscriptaddress,preprintnumbers,11pt,nofootinbib]{revtex4-1}
\pdfoutput=1
\usepackage{psfrag,graphicx,epsfig,amsmath,amssymb,bm,tikz}
\usetikzlibrary{arrows,decorations.markings,decorations.pathmorphing}

\makeatletter
\DeclareRobustCommand*{\bfseries}{%
  \not@math@alphabet\bfseries\mathbf
  \fontseries\bfdefault\selectfont
  \boldmath
}
\makeatother

\newcommand{\Lag}{\mathcal{L}}

\renewcommand{\Re}{\mathrm{Re}}
\renewcommand{\Im}{\mathrm{Im}}

\usepackage{amsmath,bbm,amssymb,slashed}
\usepackage{tikz}

\usetikzlibrary{arrows,decorations.markings,decorations.pathmorphing}

\usepackage[T1]{fontenc}
\usepackage[utf8]{inputenc}
\usepackage[english]{babel}
\usepackage{blindtext}

\usepackage[bookmarks=false]{hyperref}

\begin{document}
\preprint{FERMILAB-PUB-13-302-T}
\preprint{EFI-13-18}
\preprint{NSF-KITP-13-143}

\title{Exotic Leptons: Higgs, Flavor and Collider Phenomenology}

\author{Wolfgang Altmannshofer}
\affiliation{Fermi National Accelerator Laboratory, P.O. Box 500, Batavia, IL
60510, USA}

\author{Martin Bauer}
\affiliation{Fermi National Accelerator Laboratory, P.O. Box 500, Batavia, IL
60510, USA}
\affiliation{Enrico Fermi Institute, University of Chicago, Chicago, IL 60637,
USA}

\author{Marcela Carena}
\affiliation{Fermi National Accelerator Laboratory, P.O. Box 500, Batavia, IL
60510, USA}
\affiliation{Enrico Fermi Institute, University of Chicago, Chicago, IL 60637,
USA}
\affiliation{Kavli Institute for Cosmological Physics, University of Chicago,
Chicago, IL 60637, USA}


\begin{abstract}
\noindent
We study extensions of the standard model by one generation of vector-like
leptons with non-standard hypercharges, which allow for a sizable modification 
of the $h\to \gamma\gamma$ decay rate for new lepton masses in the 300 GeV - 1 TeV range. We analyze vacuum stability implications for different hypercharges.
Effects in $h \to Z \gamma$ are typically much smaller than in 
$h\to\gamma\gamma$, but distinct among the considered hypercharge 
assignments.
Non-standard hypercharges constrain or entirely forbid possible 
mixing operators with standard model leptons. As a consequence, the
leading contributions to the experimentally strongly constrained electric 
dipole moments 
of standard model fermions are only generated
at the two loop level by the new CP violating sources of the considered setups. 
We derive the bounds from dipole moments,
electro-weak 
precision observables and
lepton flavor violating processes, and discuss their implications. Finally, we
examine the production and 
decay channels of the vector-like leptons at the LHC, and find that signatures with multiple light leptons 
or taus are already probing interesting 
regions of parameter space.
\end{abstract}

\maketitle

\section{Introduction}

In 2012, the Large Hadron Collider 
(LHC) experiments ATLAS and CMS both
reported the discovery of a
Higgs-like boson at $m_h\simeq125$ GeV~\cite{Aad:2012tfa,Chatrchyan:2012ufa}. 
At this mass, many of its Standard Model
(SM) couplings are experimentally accessible, which makes
the question whether new physics 
manifests itself through modifications of these couplings one of the most 
interesting ones for the LHC to answer. The
hints for a possible enhancement of the Higgs to diphoton
decay rate~\cite{ATLAS:2012ad,ATLAShgg2012,Chatrchyan:2012twa,CMShgg2012} 
therefore triggered a lot of activity in the model building
community (see e.g.
\cite{Carena:2011aa,Ellwanger:2011aa,Batell:2011pz,Arhrib:2011vc,Arhrib:2012ia,
Wang:2012zv,Akeroyd:2012ms,Carena:2012xa,An:2012vp,Joglekar:2012vc,
ArkaniHamed:2012kq,
Almeida:2012bq,Delgado:2012sm,Kearney:2012zi,SchmidtHoberg:2012yy,
Davoudiasl:2012ig,McKeen:2012av,Wang:2012ts,Bertuzzo:2012bt,
Batell:2012ca,Altmannshofer:2012ar,Moreau:2012da,Batell:2012zw,
Davoudiasl:2012tu,Kopp:2013mi,Fan:2013qn,Dev:2013ff,
Carmona:2013cq,Feng:2013mea,Joglekar:2013zya,Maru:2013ooa}). 
In the latest analysis by ATLAS
\cite{ATLAShgg2013}, this excess of events in the $h\to \gamma\gamma$
channel leads to a best fit value of the signal strength of
$1.65\pm 0.24 ^{+0.25}_{-0.18}$ times the predicted SM value, while with the 
full 7+8 TeV data set, the CMS $h\to\gamma\gamma$ signal has gone down to 
$0.78^{+0.28}_{-0.26}$~\cite{CMShgg2013}.
The $h\to \gamma\gamma$ decay is loop induced in the SM and therefore highly 
sensitive to new physics effects. It will be extremely interesting to monitor 
if a significant deviation from the SM prediction can be established at the 13 
TeV LHC run.
Although all other presently measured decay rates are compatible
with the SM predictions, there is still room to consider NP effects, 
and it is also interesting to explore connections with modifications 
in these channels. Here we concentrate on the properties of models which 
can lead to a significant modification in $h\to \gamma\gamma$, without
sizable effects in other channels. The properties of such models can be 
narrowed down
considerably~\cite{Carena:2012xa}.

A promising class of models to this end are extensions of the SM by a set of
new vector-like leptons, transforming as electro-weak doublets and singlets,
respectively. Sizable Yukawa couplings between the Higgs and these new
states allow for a modification of the $h\to\gamma\gamma$ rate without 
modifying the main production process via gluon fusion. The price for
this modification is a severe vacuum instability bound, because the new
Yukawa couplings will drive the Higgs quartic coupling negative at a scale
around $10$ TeV~\cite{ArkaniHamed:2012kq,Joglekar:2012vc,Reece:2012gi}. In 
addition,
possible mixing terms between the new vector-like leptons and the SM leptons 
can induce
1-loop contributions to electric and magnetic dipole moments (EDM/MDM), as
well as tree level contributions to lepton flavor violating processes, which 
are strongly constrained
experimentally~\cite{McKeen:2012av,Fan:2013qn,Ishiwata:2013gma}. 
While the vacuum instability bound calls for an 
extension of this model at a relatively low scale (see 
e.g.~\cite{Batell:2012zw,Joglekar:2013zya,Dermisek:2013gta}), the latter 
constraints have
been usually avoided in the literature by either assuming very small 
coefficients
for the mixing operators or by invoking a discrete symmetry
\cite{ArkaniHamed:2012kq,Joglekar:2012vc,Davoudiasl:2012ig}. In this work, we 
argue that a different 
hypercharge
assignment to the new vector-like leptons can in principle not only relax the 
vacuum 
instability bound,
but simultaneously also ensures automatically that the leading contributions to 
dipole
moments only arise at the 2-loop level.

We explore two scenarios: One, in which the hypercharge of the weak doublets of 
new
vector-like leptons is $Y=-3/2$ and one where it is $Y=-5/2$. In
Section~\ref{models},
we introduce these models and discuss the masses and couplings of the new 
leptons.
One important difference between
the two scenarios is, that the first model allows for a single renormalizable
coupling to SM leptons,
while in the latter only non-renormalizable operators can couple the
vector-like leptons to the SM leptons.
In Section~\ref{hgaga}, we discuss the modified Higgs phenomenology of the 
models. 
We compute the $h \to \gamma \gamma$ and $h \to Z 
\gamma$ decay rates, which are both affected by the presence of relatively 
light vector-like leptons. We also discuss the vacuum stability bounds in this 
setup. 
In Section~\ref{EDMMDM}, we discuss the
implications of the new vector leptons for EDMs, MDMs, as well as the S and T 
parameter.  
In Section~\ref{constraints}, $Z$ pole observables and lepton flavor 
violating processes are used to derive bounds on the
model parameters that mix the vector-like leptons with SM leptons. 
Such constraints are also 
highly relevant for 
scenarios in which the new vector leptons share the quantum
numbers with the SM leptons.
Finally, in Section~\ref{secdecay}, we
discuss the production and decay processes of the new vector leptons. We 
compute the 
relevant cross
sections of final states with multiple light leptons and taus
and confront them with the existing 
searches at the LHC.

\section{Models with Exotic Hypercharges}\label{models}
\begin{figure}[tb]
 \begin{center}
\begin{tabular}{cc}
\begin{tikzpicture}[rotate=180,baseline=0,scale=1.2]
\draw[decoration={coil,aspect=0,segment
length=2.5mm,amplitude=.7mm},thick,decorate] (-2,.8) to (-1,.8);
\draw[decoration={coil,aspect=0,segment
length=2.5mm,amplitude=.7mm},thick,decorate] (-2,-.8) to (-1,-.8);
\draw[thick] (0,0) to (-1,0.8);
\draw[thick]  (0,0) to (-1,-.80);
\draw[decoration={coil,aspect=0,segment
length=2.5mm,amplitude=.7mm},thick,dashed]  (0,0) to (1,0);
\draw[thick] (-1,-0.8) to (-1,0.8);
\draw[thick] (-1,-0.8) to (-1,0.8);
\node[] at (.7,-0.3) {$h$};
\node[] at (-2.2,1) {$\gamma$};
\node[] at (-2.2,-1) {$\gamma$};
\fill[color=black!] (0,0) circle (0.3ex);
\fill[color=black!] (-1,-0.8) circle (0.3ex);
\fill[color=black!] (-1,0.8) circle (0.3ex);
\end{tikzpicture}\hspace{2cm}&\hspace{2cm}
 \begin{tikzpicture}[scale=1.2,baseline=0]
\draw[thick,dashed] (-0.5,1) to (0.5,1);
\draw[thick,dashed] (-0.5,-1) to (0.5,-1);
\draw[thick] (0.5,1) to (2.5,1);
\draw[thick,dashed] (2.5,1) to (3.5,1);
\draw[thick,dashed] (2.5,-1) to (3.5,-1);
\draw[thick] (0.5,-1) to (2.5,-1);
\fill[color=black!] (0.5,1) circle (0.3ex);
\fill[color=black!] (0.5,-1) circle (0.3ex);
\fill[color=black!] (2.5,1) circle (0.3ex);
\fill[color=black!] (2.5,-1) circle (0.3ex);
\draw[ thick] (2.5,-1) to (2.5,1);
\draw[ thick] (0.5,-1) to (0.5,1);
\node[] at (-.2,-0.7) {$h$};
\node[] at (-.2,0.7) {$h$};
\node[] at (3.2,-0.7) {$h$};
\node[] at (3.2,0.7) {$h$};
\end{tikzpicture}
\end{tabular}
\end{center}
\caption{
\label{fig:FeynmanDiagram} \small
Feynman diagrams involving the new vector-like leptons (straight lines), which
are responsible for 1-loop contributions to the $h\to\gamma\gamma$ decay rate
(left) and the running to the Higgs quartic coupling (right). 
}
\end{figure}
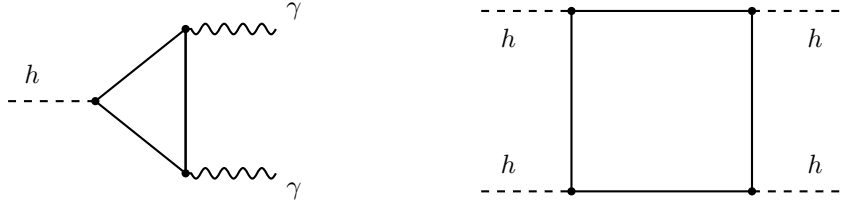

\noindent
The loop induced coupling of the Higgs to two photons is sensitive to the Yukawa
coupling of any internal fermion and to its electric charge, see the diagram on
the left hand side of Figure~\ref{fig:FeynmanDiagram}. 
As is well known~\cite{Carena:2012xa}, the contributions from chiral fermions 
interfere destructively with the $W$ loop contribution, which dominates in the 
SM.
On the other hand, extensions of the SM with vector-like leptons which couple 
to the Higgs, can both enhance and suppress the $h\to \gamma\gamma$ decay rate, 
because in contrast to chiral leptons, vector-like leptons can interfere also 
constructively with the $W$ contribution.
Colored new matter is additionally constrained by data, because it also affects
the gluon fusion production cross section of the Higgs, which in turn leads to 
modifications of all Higgs signal strengths~\cite{Dobrescu:2011aa,
Carena:2011aa,Batell:2011pz,
Carena:2012xa,Batell:2012ca,Moreau:2012da,Kumar:2012ww,Dawson:2012di,
Buckley:2012em,Bertuzzo:2012bt,Fajfer:2013wca}. 
 
A sizable correction of the $h \to \gamma\gamma$ rate due to vector-like 
leptons, requires large Yukawa
couplings of the leptons to the
Higgs, which in turn contribute to the beta function of the Higgs quartic
coupling via the box diagrams depicted on the right-hand side of 
Figure~\ref{fig:FeynmanDiagram}. This will cause the Higgs quartic coupling to 
run negative at a very low scale \cite{ArkaniHamed:2012kq}.
We argue, that this scale can be in principle considerably 
higher, if the 
new vector-like
leptons carry larger electric charges. Larger electric charges affect the $h\to 
\gamma\gamma$
decay rate, but are not felt by the Higgs quartic coupling at the 1-loop level.

Therefore, we study the effects of 
extending the SM by one generation of new vector leptons, which transform
as doublets and singlets under
$SU(2)_L\times U(1)_Y$, 
\begin{align}
 L_L &= \begin{pmatrix}N_L \\ E_L \end{pmatrix} = (1,2)_{-1/2-n} ~~,
\hspace{2cm} \tilde L_R
= \begin{pmatrix} \tilde N_R \\ \tilde E_R \end{pmatrix} = (1,2)_{-1/2-n} ~,
\notag\\
 \tilde E_L &= (1,1)_{-1-n} ~~, \hspace{3.9cm}E_R = (1,1)_{-1-n} ~\,.
\end{align}
The case $n=0$ corresponds to the widely
discussed scenario in which the quantum numbers are a copy of the ones of the SM
leptons 
\cite{ArkaniHamed:2012kq,Joglekar:2012vc,Kearney:2012zi,Davoudiasl:2012ig,
Voloshin:2012tv,McKeen:2012av,Fan:2013qn,Garg:2013rba}. In addition to a strong vacuum 
stability constraint, the $n=0$ setup allows for direct mass mixing
operators with the SM leptons.
This can lead to large corrections of the couplings of the SM leptons with the 
weak gauge bosons and the Higgs and also induce tree level FCNC couplings. 
In addition, large 
1-loop contributions to EDMs and MDMs can be generated. The measurements of 
EDMs, 
MDMs, lepton flavor violating observables and $Z$ pole observables therefore 
constrain the coefficients of these mass mixing operators to be extremely 
small. One way to
account for these bounds is to invoke a discrete symmetry which forbids
mixing operators \cite{Joglekar:2012vc}. In the models considered in this
article however, 
these operators are either absent in the $n=2$ case, due to the
different
hypercharges between the vector-like and the SM leptons, 
or only one operator is allowed that couples the new sector 
just with the right handed SM sector, limiting the impact of 
the mass mixing operators for $n=1$.
In particular for $n=1$, we find the following mass
terms and Yukawa couplings,
\begin{eqnarray}\label{deltalag}
 -\Delta\Lag &=& \,\, M_L \bar L_L \tilde L_R + M_E \bar{\tilde{E}}_L E_R + 
y_E\,\bar{\tilde{L}}_R \,h\, \tilde E_L +  y_L \bar L_L \,h\, E_R 
\nonumber \\
&&+ 
\sum_{\ell=e,\mu,\tau}\,y^{\rm SM}_\ell\,\bar \ell_L \,h\, \ell_R +
 y_{L\ell}\, \bar L_L \,\tilde h\, \ell_R +h.c. ~,
\end{eqnarray}
in which SM fields are denoted by lower case letters, and $\tilde h \equiv
i\sigma_2
h^\dagger$.
Note that the hypercharge assignment only allows one mixing operator, that 
mixes 
the right-handed SM leptons with the left-handed doublet of vector leptons.
The Lagrangian for the general scenario $n>1$ corresponds to
\eqref{deltalag} with the coefficient of the mixing operator set to zero,
$y_{L\ell}=0$.
We only discuss the scenarios
$n=1$ and $n=2$ here. Models with even higher hypercharges
might lead to interesting phenomenology as well, but result in a 
Landau pole of the
hypercharge gauge coupling at scales of $\sim 10^4$~TeV or below, 
see Section~\ref{sec:vacuum}.

For $n=2$, the new leptons carry electric charges $Q_E = -3$ and $Q_N = -2$.
After
electro-weak symmetry breaking (EWSB), the Lagrangian \eqref{deltalag} 
leads to the following mass terms
\begin{align}\label{lag1}
-\Delta\Lag \supset (\bar E_L, \bar{\tilde{E_L}}) \bm{\mathcal{M}}_E 
\begin{pmatrix} \tilde E_R \\ E_R \end{pmatrix} + \bar N_L \bm{\mathcal{M}}_N 
\tilde N_R\, + m_\ell \bar \ell_L\ell_R ~~+ h.c.~,
\end{align}
with the masses of the SM leptons $m_\ell = y_\ell^\text{SM} v$ and the Higgs 
vev, $v= 174$ GeV.
In the absence of a doubly charged singlet, the mass of the charge two 
component of $L$
will only be given by its vector mass, while the charge three leptons mix,
\begin{align}\label{n=2mass}
 n=2:\qquad \bm{\mathcal{M}}_E =\begin{pmatrix}
                    M_L & v\,y_L\\
                    v\,y_E^\ast& M_E
                   \end{pmatrix}\,,\qquad  \bm{\mathcal{M}}_N=M_L\,.
\end{align}
While the parameters $M_L$, $M_E$, $y_L$ and $y_E$ can all be complex, three 
phases can be absorbed by re-phasing the vector lepton fields, leaving one 
physical CP violating phase $\tilde \phi = \text{Arg}(M_L M_E y_L^* y_E)$. In 
the following we will work in a convention where the vector masses are real and 
positive and parametrize the physical phase by the relative phase of the 
Yukawa couplings, {\it i.e.} $\tilde \phi = \text{Arg}(y_L^* y_E)$.  

For the case of $n=1$, the new leptons carry electric charges $Q_E=-2$ and
$Q_N=-1$, and the mass Lagrangian reads
\begin{align}
-\Delta\Lag \supset (\bar E_L, \bar{\tilde{E_L}}) \bm{\mathcal{M}}_E 
\begin{pmatrix} \tilde E_R \\ E_R \end{pmatrix} + (\bar\ell_L, \bar N_L)\,
\bm{\mathcal{M}}_{N} \,\begin{pmatrix}
                \ell_R  \\ \tilde N_R 
                \end{pmatrix} ~~+ h.c.~,
\end{align}
Mixing with the SM leptons is generated proportional to the Yukawa couplings 
$y_{L\ell}$, so that  
\begin{align}\label{n=1mass}
 n=1:\qquad\bm{\mathcal{M}}_E =\begin{pmatrix}
                    M_L & v\,y_L\\
                    v\,y_E^\ast& M_E
                   \end{pmatrix}\,,\qquad \bm{\mathcal{M}}_N =\begin{pmatrix}
                    v\,y^\mathrm{SM}_\ell  & 0\\
                    v\,y_{L\ell}& M_L
                   \end{pmatrix}\,,
\end{align}
in which only mixing with one SM lepton generation is considered for 
simplicity. The extension to the 3 generation case is straightforward. The 
phases of the mixing Yukawas $y_{L\ell}$ are additional physical sources of CP 
violation.

Both in the $n=1$ and $n=2$ case, we can diagonalize the mass matrix 
$\bm{\mathcal{M}}_E$ by a bi-unitary transformation $Z_L \bm{\mathcal{M}}_E 
Z_R^\dagger = \text{diag}(m_1, m_2)$ and we introduce two Dirac spinors 
$(\chi_1, \chi_2)^T = Z_L (E_L, \tilde E_L)^T + Z_R (\tilde E_R, E_R)^T$ to 
describe the light and heavy mass eigenstates with masses
\begin{align}
 m_{1,2}^2=\frac{1}{2}\Big(M_L^2+&M_E^2+v^2(|y_E|^2+|y_L|^2)\notag\\
 \mp&\sqrt{
\big(M_L^2+M_E^2+v^2(|y_E|^2+|y_L|^2)\big)^2 - 4\big| M_L M_E - v^2 y_L y_E^* 
\big|^2 }\Big) ~.
\end{align}

In the $n=2$ scenario, the charge two lepton $N = N_L + \tilde N_R$ is its own 
mass
eigenstate with $m_N = M_L$. In the $n=1$ scenario however, this is only true 
up to corrections
proportional to the mixing coefficients $y_{L\ell}$. As we will see in the 
following, the size of the mixing between the SM leptons and the vector leptons 
is constrained to be small. We therefore treat this mixing perturbatively and 
find the following leading corrections to the masses of $N$ and the SM leptons  
\begin{equation}
 m_N = M_L\left(1+\frac{1}{2} \sum_\ell 
|y_{L\ell}|^2\,\frac{v^2}{M_L^2}+\ldots\right) 
~,~~~ m_\ell =
v y_\ell^\text{SM} 
\left(1-\frac{1}{2}|y_{L\ell}|^2\,\frac{v^2}{M_L^2}+\ldots\right)\,.
\end{equation}
In the $n=1$ case, the mixing terms also lead to modifications of the couplings 
of the SM leptons, once one rotates into the mass eigenstate basis.
In particular, the flavor diagonal couplings of the Higgs to SM leptons and of 
the $Z$ boson to right-handed SM leptons are modified at the order 
$|y_{L\ell}|^2 
v^2/M_L^2$. Moreover, flavor changing couplings of the Higgs to SM leptons and 
of the $Z$ boson to right-handed SM leptons are induced at the same order, 
provided that the new leptons couple to 
at least two families of SM leptons simultaneously, see e.g. 
eqs.~(\ref{ghetaLetaR}) and (\ref{gZetaRetaR}) in Appendix~\ref{coups}. In 
addition, couplings of the $W$ boson to the new charge one leptons and SM 
neutrinos as well as to the new charge 2 leptons and SM leptons are generated.
Explicit expressions for all the couplings that are relevant for our analysis 
are collected in the Appendix~\ref{coups}.

\section{\boldmath{$h\to \gamma\gamma$}, \boldmath{$h\to Z\gamma$}, and Vacuum 
Instability Constraints} \label{hgaga}

\subsection{The $h\to \gamma\gamma$ Rate}

Based on the mass matrices given in eqns. \eqref{n=2mass} and
\eqref{n=1mass}, one can obtain the contribution to the 
$h\to \gamma\gamma$ decay rate at leading order in the electro-weak scale 
over the vector masses, using low energy 
theorems~\cite{Ellis:1975ap,Shifman:1979eb,Voloshin:2012tv}. 
Notice, that in contrast to the $n=0$ scenario, for $n=1$ 
there is only one off-diagonal
mixing term between the vector-like and the SM leptons. As a consequence,
both in the $n=2$ and $n=1$ scenarios, at leading order, the only 
non-SM contributions to the $h \to \gamma \gamma$ decay rate 
are generated by the mass matrix $\bm{\mathcal{M}}_E$. 
In contrast to the case of chiral fermions, the effective interaction of the 
Higgs with photons contains both a CP-even and a CP-odd 
part~\cite{Voloshin:2012tv}
\begin{equation}
 \mathcal{L} \supset \frac{\alpha_\text{em}}{4 \pi} Q_\chi^2 \frac{h}{v} \left( 
\frac{1}{3} F_{\mu\nu} F^{\mu\nu} \frac{\partial}{\partial \log v} \log \det 
\left( \bm{\mathcal{M}}_E^\dagger \bm{\mathcal{M}}_E \right) + \frac{1}{2} 
\epsilon^{\mu\nu\rho\sigma} F_{\mu\nu} F_{\rho\sigma} \frac{\partial}{\partial 
\log v} \arg \det \bm{\mathcal{M}}_E \right) ~.
\end{equation}
As corrections to the Higgs
production cross section are negligible in our framework, the ratio of the 
Higgs diphoton rate normalized to the respective SM rate is to an excellent 
approximation given by the ratio of the $h\to \gamma\gamma$ partial decay 
widths. We find
\begin{align}
 R_{\gamma\gamma} & 
=\frac{\sigma(p p \to h)}{\sigma_\mathrm{SM}(p p\to h)}
\frac{\Gamma(h\to \gamma\gamma)}{\Gamma_\mathrm{SM}(h\to\gamma\gamma)}
\approx \frac{\Gamma(h\to \gamma\gamma)}{\Gamma_\mathrm{SM}(h\to \gamma\gamma)} 
\notag \\
\label{Rgg}
 & \approx \left\vert1+Q_\chi^2 \frac{4}{3} \frac{v\,\partial_v \log (\det
\bm{\mathcal{M}}_E^\dagger \bm{\mathcal{M}}_E)}{A_1(\tau_W)+\frac{4}{3}
A_{1/2}(\tau_t)}\right\vert^2+\left\vert
4 Q_\chi^2\frac{v\,\partial_v \arg (\det 
\bm{\mathcal{M}}_E)}{A_1(\tau_W)+\frac{4}{3} A_{1/2}(\tau_t)}\right\vert^2 ~ ,
\end{align}
which is valid for both scenarios.
Here, $\tau_i = 4 m_i^2/m_h^2$ and we neglected the tiny bottom quark 
contribution 
to the SM width. To a good approximation, one has for the SM $W$ and top loops
$A_1(\tau_W)+\frac{4}{3} 
A_{1/2}(\tau_t) \approx -8.3 + 1.8 \approx -6.5$. Expressions for the
loop functions $A_1$ and $A_{1/2}$
are collected in Appendix~\ref{loopfuncs}.
The explicit form of the derivatives of the mass matrix read
\begin{align} \label{CPeven}
 v\,\partial_v \log (\det
\bm{\mathcal{M}}_E^\dagger \bm{\mathcal{M}}_E) & = -4 ~\frac{M_EM_L 
\Re{(y_L^\ast y_E)}v^2-|y_E
y_L|^2\,v^4}{\vert M_EM_L-y_E^\ast y_L\,v^2\vert^2} \simeq 
-4\frac{|y_L||y_E|v^2}{M_E M_L} \cos\tilde\phi~, \\ \label{CPodd}
v\,\partial_v \arg (\det \bm{\mathcal{M}}_E) & = -2 ~\frac{M_L M_E 
\Im{(y_L^\ast y_E})v^2}{\vert
M_EM_L-y_E^\ast y_L\,v^2\vert^2} \simeq -2\frac{|y_L| |y_E| v^2}{M_E 
M_L} \sin\tilde\phi ~.
\end{align}
Note that the CP-odd contribution does not interfere with the SM amplitude. 
Even though the CP-odd part will therefore always
enhance the $h\to\gamma\gamma$ cross section, it will typically amount to 
at most a
percent correction for all phenomenologically viable parameters of the 
considered model. In 
Section~\ref{EDMMDM}, we will see that even this is very optimistic, given the 
very
stringent bounds on the new physics phase, $\tilde\phi$, coming from the 
electron EDM. The 
CP-even
part in \eqref{Rgg} interferes with the SM contribution and therefore allows for
significantly larger corrections. Depending on the overall sign of the
numerator in~(\ref{CPeven}), this can lead to an enhancement or decrease of the 
$h\to\gamma\gamma$ cross section.
The term $\sim |y_Ey_L|^2$ in~(\ref{CPeven}) always leads to a decreased
cross section compared to the
SM, but can be neglected to a first approximation. The term could only become 
relevant for very small vector masses $M_L$ and $M_E$, that are strongly 
constrained by direct searches, or for large Yukawa couplings, that are 
theoretically unattractive, as they imply large corrections to the running of 
the Higgs quartic coupling, forcing it to
become negative at very low scales.
The sign of the interference is therefore mainly determined by the sign of 
Re$(y_L^\ast y_E) = |y_L| |y_E| \cos\tilde\phi$.

While the expression~(\ref{Rgg}) captures the leading contributions in an 
expansion in the ratio of the electro-weak scale over the vector masses, one 
can easily go beyond this approximation working with mass eigenstates of the 
new leptons.
Doing so, the corrections to the Higgs diphoton decay rate can be written as
\begin{align} \label{eq:exact}
R_{\gamma\gamma} = \left\vert1+Q_\chi^2 \frac{\sum_i \frac{v}{m_i} 
\text{Re}(g_{h\chi_i\chi_i}) A_{1/2}(\tau_{\chi_i})}{A_1(\tau_W)+\frac{4}{3}
A_{1/2}(\tau_t)}\right\vert^2+\left\vert
Q_\chi^2 \frac{\sum_i \frac{v}{m_i} \text{Im}(g_{h\chi_i\chi_i}) \tilde 
A_{1/2}(\tau_{\chi_i})}{A_1(\tau_W)+\frac{4}{3} A_{1/2}(\tau_t)}\right\vert^2 ~.
\end{align}
The expressions for the couplings $g_{h\chi_i\chi_i}$ of the Higgs with the new 
lepton mass eigenstates are given in Appendix~\ref{coups}.
In the $n=1$ case, there are in principle also contributions from the charge 1 
leptons that are formally of higher order in $v^2/M_L^2$. Working with mass 
eigenstates, they can be taken into account in a straight forward way. 
However, given the constraints on the mixing Yukawas that will be discussed in 
Section~\ref{constraints}, we find that contributions from the new charge 1 
states 
are negligible even for very light masses $M_L =O(v)$.
Expanding~(\ref{eq:exact}) in $v/M$ we recover the approximate expression 
in~(\ref{Rgg}).
We find that~(\ref{Rgg}) is accurate at the one percent level as long as the 
vector masses are $M_L, M_E \gtrsim 300$~GeV.
In our numerical analysis, we work with mass eigenstates, though. 

Due to their large charges, the new leptons can lead to sizable effects in $h 
\to \gamma\gamma$ even for moderate values of the Yukawa couplings. 
In particular, for fixed
vector masses $M_L$ and $M_E$, the Yukawa couplings can be smaller by a
factor 
\begin{align}
 y\,\to\, \frac{y}{Q_\chi}\,,
\end{align}
while keeping the decay rate constant compared to the $Q_\chi=1$~$(n=0)$ 
scenario. Conversely, for fixed Yukawa couplings, higher charges allow for 
heavier vector masses.

\subsection{Vacuum Stability} \label{sec:vacuum}

\begin{figure}[tb]
\begin{center}\hspace{-10mm}
\includegraphics[width=0.5\textwidth]{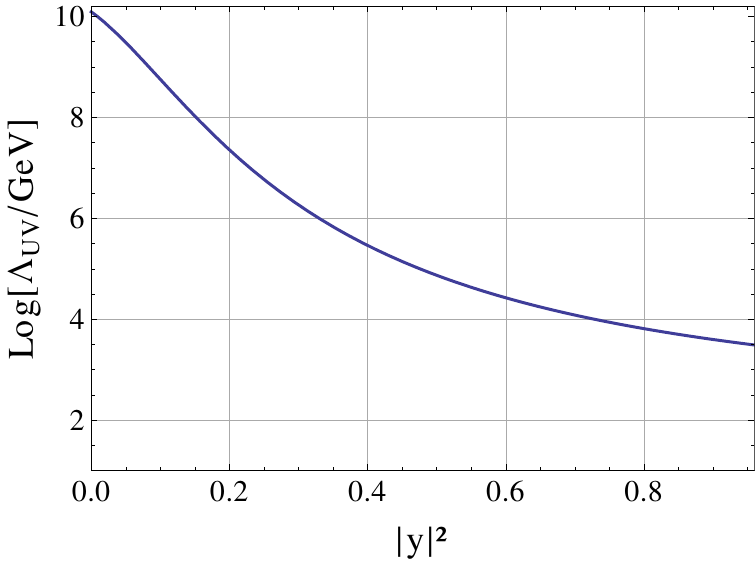}
\end{center}
\vspace{-5mm}
\caption{\small
Scale at which the scalar quartic coupling turns 
negative due to renormalization group running, as function of the Yukawa 
couplings $|y|=|y_E|=|y_L|$.}
\label{fig:Plot1b}
\end{figure}

\begin{figure}[tb]
\begin{center}\hspace{-10mm}
\begin{tabular}{lr}
\includegraphics[width=0.5\textwidth]{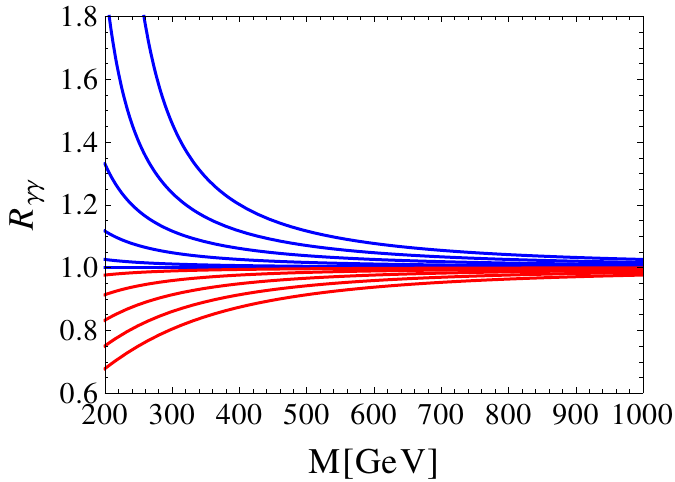}
&\includegraphics[width=0.5\textwidth]{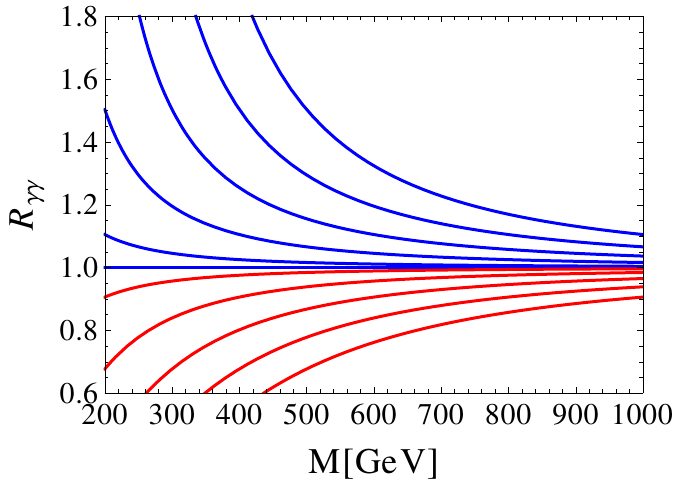}
\end{tabular}
\includegraphics[width=0.5\textwidth]{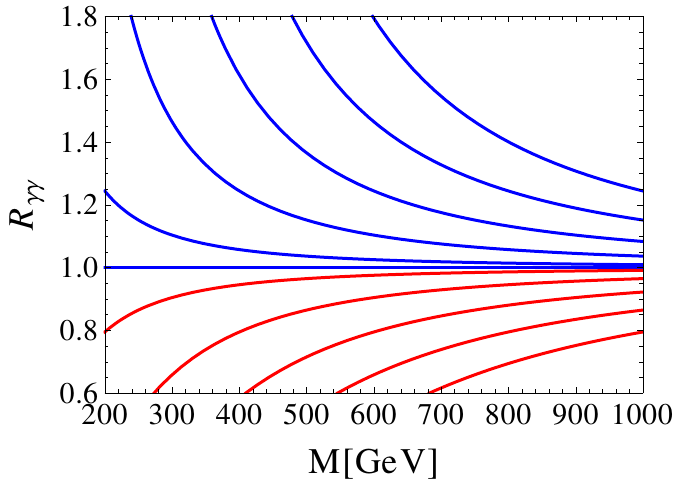}
\end{center}
\vspace{-5mm}
\caption{\small
Modifications of the diphoton decay rate of the Higgs versus the vector mass
$M=\sqrt{|M_E M_L|}$ for $n=0$ and $n=1$ on the upper left and right panel, as 
well 
as for $n=2$ in the lower panel. The blue curves are for $\sqrt{y_E y_L} =\{0, 
0.2, 0.4, 
0.6, 0.8, 1\}$, and the red curves correspond to $\sqrt{- y_E y_L} = \{0.2, 
0.4, 0.6, 0.8, 
1\}$. For all three plots, $y_E=y_L=0$ corresponds to $R_{\gamma\gamma}=1$ and 
the effects become larger for larger absolute values of $y_E$ and $y_L$.}
\label{fig:Plot1}
\end{figure}

The existence of Yukawa type interactions, with order one couplings,  of the 
Higgs with the new leptons has important implications for the stability of 
the Higgs potential.
The Yukawa couplings contribute at 1-loop to the running of the Higgs quartic
coupling through the box diagram on the right hand side of 
Figure~\ref{fig:FeynmanDiagram}. We find a correction to the SM beta function 
of\footnote{In the n=1 case there are additional contributions to the beta 
function coming from the mixing Yukawas $y_{L\ell}$. In regions of parameter 
space where the vector leptons can lead to sizable modifications of the 
$h\to\gamma\gamma$ rate, they are bound to be 
small from indirect constraints (see Section~\ref{constraints}). Therefore 
their impact on 
the running of the Higgs quartic is negligible.}
\begin{equation} \label{eq:betalambda}
 \frac{d \lambda}{dt} = \frac{1}{16\pi^2} \beta_\lambda = \frac{1}{16\pi^2} 
\Big( \beta_\lambda^{\rm SM} + 4 \lambda(y_E^2 + y_L^2) - 4(y_E^4 + y_{L}^4) 
\Big) ~.
\end{equation}
The scale at which the quartic coupling runs negative is plotted in Figure 
\ref{fig:Plot1b} versus the absolute value of the new Yukawa couplings 
$|y|=|y_E|=|y_L|$.
For $y=0$ one recovers the SM limit that, for values of 
$\alpha_s = 0.1184$~\cite{Bethke:2009jm} and $m_t = 
173.2$~GeV~\cite{Lancaster:2011wr}, and considering a two loop renormalization 
group running, yield a vanishing value of $\lambda$ at a UV scale of 
$\Lambda_\text{UV} \simeq 10^{10}$~GeV 
(see e.g.~\cite{Degrassi:2012ry,Buttazzo:2013uya}). 

The effect of non-zero Yukawa couplings $|y|=|y_E|=|y_L|$ on the vacuum 
stability of 
the Higgs potential has to be compared with the effects in 
$R_{\gamma\gamma}$, 
which are shown in Figure~\ref{fig:Plot1} as a function of 
the geometric mean of the vector masses $M=\sqrt{M_E M_L}$ for $n=1$ in the 
upper right 
panel and for
$n=2$ in the lower panel. The $n=0$ scenario is shown in the upper left 
panel for comparison. Each blue curve corresponds, from bottom to top, to 
Yukawa 
couplings $\sqrt{y_E y_L} = \{ 0, 0.2, 0.4, 0.6, 0.8, 1\}$, and each red 
curve 
corresponds, from top to bottom, to $\sqrt{- y_E y_L} = \{ 0.2, 0.4, 0.6, 0.8, 
1\}$. For $y=0$ 
one has 
$R_{\gamma\gamma}=1$ and the effects are larger for larger absolute values of 
$y$. A possible phase $\tilde \phi = \text{Arg}(y_L^\ast y_E)$ is set to zero 
in the plots.

If one requires an enhancement of the Higgs diphoton rate by 30\%, one finds in
the $n=0$ case that the Higgs quartic coupling runs negative at
$\Lambda_\mathrm{UV}\approx 10-100$ TeV, even for the most optimistic 
assumptions 
like the lightest mass eigenstate close to the LEP bound $m \sim 100$~GeV~,  
in agreement with Ref.~\cite{ArkaniHamed:2012kq}.
As a consequence, such models would require a UV completion at or below
the 10-100~TeV scale.
This bound can in principle be relaxed considerably for
the scenarios considered in this work.
For the same spectrum and the same enhancement of the Higgs diphoton rate, the 
scales where the Higgs quartic runs negative can 
be as high as $10^5$~TeV in the $n=1$ case and $10^6$~TeV in the $n=2$ case.\footnote{A comparable suppression of the Higgs di-photon rate requires slightly larger Yukawa couplings and therefore slightly smaller UV scales.}
However, present LHC searches in multilepton channels, including taus, 
can already start probing the existence of
new vector leptons.
In the analyses presented in Section~\ref{secdecay}, we show that in the 
minimal models considered here, new vector leptons are
viable if their vector masses are of the order of $M \gtrsim 370$~GeV 
in the case $n=1$ and $M \gtrsim 850$~GeV in the case $n=2$.
Therefore, given that the vector leptons have to be considerably heavy 
in the minimal setups we have investigated, it turns out that  
the UV scale where the quartic Higgs coupling becomes negative
is actually comparable to the $n=0$ case, namely around $\sim 10-100$~TeV 
in the $n=1$ case and even lower in the $n=2$ case.
As will be discussed in Section~\ref{secdecay}, in extensions of the setups 
with an additional massive neutral state and with additional 
interactions parametrized by higher dimensional operators, lighter vector-like 
leptons can become viable also for 
$n=1$ and $n=2$.

For the numerical calculation of the running we take into account the Higgs 
quartic, the SM gauge couplings, the top Yukawa and the contributions from the 
new Yukawas $y_E$ and $y_L$.
We use 2-loop expressions for the SM beta 
functions~\cite{Machacek:1983tz,Machacek:1983fi,Machacek:1984zw,Ford:1992pn,
Luo:2002ey} and add the 1-loop contributions from the new leptons. 
The running of the Higgs quartic coupling was already given 
in~(\ref{eq:betalambda}). For the 
gauge and Yukawa couplings we find
\begin{align}
 \frac{dg_1}{dt} = \frac{g_1}{16\pi^2} \beta_1&=\frac{g_1}{16\pi^2} \left( 
\beta_1^\mathrm{SM}+\frac{8}{3}\left(\frac{1}{2}+n\right)^2 
g_1^2+\frac{4}{3}(1+n)^2 g_1^2 \right) \,,\\
 \frac{dg_2}{dt} = \frac{g_2}{16\pi^2} \beta_2&=\frac{g_2}{16\pi^2} \left( 
\beta_2^\mathrm{SM}+\frac{2}{3} g_2^2 \right)\,, \\
 \frac{dy_t}{dt} = \frac{y_t}{16\pi^2} \beta_t&=\frac{y_t}{16\pi^2} \Big( 
\beta_t^\mathrm{SM} + |y_L|^2 + |y_E|^2 \Big) \,, \\
 \frac{dy_L}{dt} = \frac{y_L}{16\pi^2} \beta_L&=\frac{y_L}{16\pi^2} \left( 
\frac{3}{2} |y_L|^2 + (3 y_t^2 + |y_L|^2 + |y_E|^2) - 
\left(\left[2n^2+3n+\frac{5}{4}\right] \frac{9}{5} g_1^2 + \frac{9}{4} 
g_2^2\right) \right) \,, \\
 \frac{dy_E}{dt} = \frac{y_E}{16\pi^2} \beta_E&=\frac{y_E}{16\pi^2} \left( 
\frac{3}{2} |y_E|^2 + (3 y_t^2 + |y_L|^2 + |y_E|^2) - 
\left(\left[2n^2+3n+\frac{5}{4}\right] \frac{9}{5} g_1^2 + \frac{9}{4} 
g_2^2\right) \right) \,.
\end{align}
The beta function of the strong gauge coupling is not affected by the new 
uncolored states and we use $SU(5)$ normalization for the weak couplings $g_1^2 
= \frac{5}{3} g^2$ and $g_2^2 = g^{\prime2}$.
To first order in the ratio of the electro-weak scale over the vector masses,
there is a direct correlation of contributions to the QED beta function and the
CP-even coupling of the Higgs to two 
photons~\cite{Carena:2012xa,Ellis:1975ap,Shifman:1979eb}. Therefore, a 
modification of
$R_{\gamma\gamma}$ is necessarily correlated with a positive contribution to 
the running of
the $SU(2)_L\times U(1)_Y$ gauge couplings.
In particular, both in the $n=1$ and $n=2$ case, the running of the hypercharge 
leads to a Landau pole below
the Planck scale, but for both scenarios, this Landau
pole is orders of
magnitude above the UV scale extracted from vacuum stability considerations in 
regions of parameter space with a sizable modification of $R_{\gamma\gamma}$. It
should be mentioned, that this is not necessarily the case for scenarios with 
new leptons
that carry even larger hypercharges. For example in the $n=3$ case, the Landau 
pole arises already at a scale of $\sim10^4$ TeV.

\subsection{The $h \to Z \gamma$ Rate}

The new vector-like leptons do not only contribute at the 1-loop level to the 
$h \to \gamma \gamma$ decay, but they also modify the $h \to Z \gamma$ rate. In 
the scenario where the new leptons have the same hypercharges as the SM 
leptons, their effect in $h \to Z \gamma$ is accidentally suppressed by $1 - 
4s_W^2 \simeq 0.08$ and $h \to Z \gamma$ is to an excellent approximation 
SM-like~\cite{Carena:2012xa}. This strong suppression does not arise for our 
non-standard hypercharge assignments, and larger effects can in principle be 
expected.
 
The corrections to the $h \to Z \gamma$ rate can be written in the following 
generic form
\begin{equation}
 R_{Z \gamma} \simeq \frac{\Gamma(h\to Z\gamma)}{\Gamma_\mathrm{SM}(h\to 
Z\gamma)} = \left| 1 + \frac{F_\text{NP}}{F_\text{SM}}\right|^2 + \left| 
\frac{\tilde F_\text{NP}}{F_\text{SM}} \right|^2 ~.
\end{equation}
Here, $F_\text{SM}$ is the SM amplitude and $F_\text{NP}$ ($\tilde 
F_\text{NP}$) is the CP conserving (CP violating) part of the NP amplitude. As 
in the case of $h \to \gamma \gamma$, the by far dominant NP contributions come 
from loops involving the charge 2 states (for $n = 1$) or the charge 3 states 
(for $n=2$), respectively. Working with mass eigenstates, we find 
\begin{eqnarray}
F_\text{SM} &=& \frac{1}{s_W c_W} \left[ M_W^2 F_W + \left( 2 - \frac{16}{3} 
s_W^2\right) m_t^2 F_t \right]  ~, \\
F_\text{NP} &=& Q_\chi \sum_{j,k} \frac{v}{m_j} F(m_j,m_k) \Big[ 
\text{Re}\big(g_{h\chi_k\chi_j}^* g_{Z \chi_k\chi_j}^L\big) + 
\text{Re}\big(g_{h\chi_j\chi_k} g_{Z \chi_k\chi_j}^R\big) \Big] ~, 
\label{eq:hZga_1} \\ \label{eq:hZga_2}
\tilde F_\text{NP} &=& Q_\chi \sum_{j,k} \frac{v}{m_j} G(m_j,m_k) \Big[ 
\text{Im}\big(g_{h\chi_j\chi_k} g_{Z \chi_k\chi_j}^R\big) - 
\text{Im}\big(g_{h\chi_k\chi_j}^* g_{Z \chi_k\chi_j}^L\big) \Big] ~.
\end{eqnarray}
In the SM amplitude, we neglected the tiny contribution from the bottom quark 
loop. The $W$ and top contributions, $F_W$ and $F_t$, can be found for example 
in~\cite{Djouadi:1996yq}. Numerically, we find approximately $M_W^2 F_W \simeq 
5.1$ and $m_t^2 F_t \simeq -0.36$. The loop functions in the NP amplitudes are 
given by
\begin{equation}
 F(m_j,m_k) = m_j^2 f(m_j,m_k,m_k) ~, ~~~  G(m_j,m_k) = m_j^2 g(m_j,m_k,m_k) ~,
\end{equation}
with $f$ and $g$ given in~\cite{Djouadi:1996yq}. The relevant couplings of the 
Higgs and the $Z$ boson to the new lepton mass eigenstates are collected in 
Appendix~\ref{coups}. Note that~(\ref{eq:hZga_1}) and~(\ref{eq:hZga_2}) contain 
contributions from loops where both mass eigenstates enter simultaneously. 
These contributions are parametrically of the same order as the contributions 
from loops that contain only one mass eigenstate. 

In order to obtain an analytical understanding of the NP contributions to $h 
\to Z \gamma$, we expand the corrections to $R_{Z\gamma}$ to leading order in 
the electro-weak scale over the vector masses. We find
\begin{eqnarray}
F_\text{NP} &=& -\frac{Q_\chi}{s_W c_W} \left[ \Big( 1 + 4 Q_\chi s_W^2 + 
h_1(x) \Big) \frac{2}{3} \frac{|y_L| |y_E| v^2}{M_L M_E} \cos\tilde\phi + 
h_2(x) 
\frac{(|y_L|^2 + |y_E|^2)v^2}{M_L^2} \right] ~, \\
 \tilde F_\text{NP} &=& -\frac{Q_\chi}{s_W c_W} \Big( 1 + 4 Q_\chi s_W^2 + 
h_3(x) \Big) \frac{1}{3} \frac{|y_L| |y_E| v^2}{M_L M_E} \sin\tilde\phi ~.
\end{eqnarray}
The functions $h_1$, $h_2$, and $h_3$ depend on the ratio of the vector masses 
$x = M_E^2 / M_L^2$ and for degenerate masses we have $h_1(1) = h_2(1) = h_3(1) 
= 0$. The explicit expressions for the $h_i$ functions are given in the 
appendix. Even for large splittings of the vector masses, we find that the 
effects of the 
$h_i$ is typically small. Therefore,
we indeed observe that in the $n=0$ case, the corrections to $h \to 
Z\gamma$ are accidentally suppressed by $1 - 4s_W^2$, while such a suppression 
is absent in the $n=1$ and $n=2$ cases.  

\begin{figure}[tb]
\begin{center}
\includegraphics[width=0.6\textwidth]{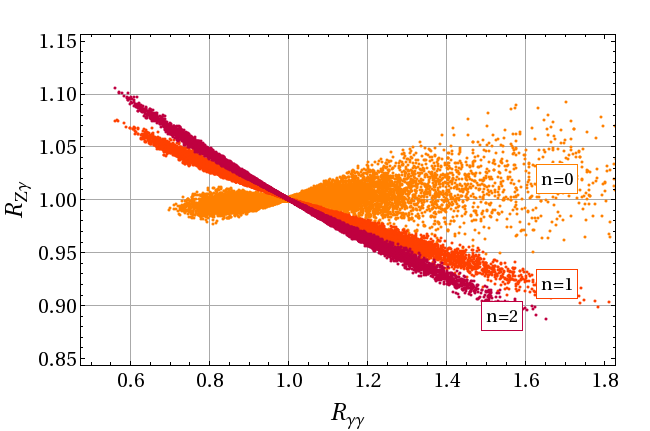}
\end{center}
\caption{\small
Correlations between the NP effects in $h \to \gamma\gamma$ and $h \to Z 
\gamma$ in the cases $n=0$ (yellow), $n=1$ (orange), and $n=2$ (purple) as 
indicated.}
\label{fig:RZga}
\end{figure}

Figure~\ref{fig:RZga} shows the correlation of NP effects in $h \to \gamma 
\gamma$ and $h \to Z \gamma$ for the 3 scenarios $n=0,1,2$. In all the cases, 
the Yukawa couplings $y_L$ and $y_E$ are taken to be real and are varied 
between~$-1$ and~$1$. The vector masses $M_L$ and $M_E$ are allowed to vary in 
the ranges $200 - 400$~GeV, $400 - 600$~GeV, and $600 - 1000$~GeV for $n=0$, 
$n=1$, and $n=2$, respectively. 
As expected, the modifications of the $h\to Z\gamma$ rate in the $n=0$ scenario 
are on average small, and can reach at most values between $-5\%$ to $+10\%$ 
for a strongly enhanced $h\to \gamma\gamma$ rate. For the scenarios with the 
larger hypercharges, the effects in $h \to Z\gamma$ can be slightly larger, but 
still 
typically do not exceed $\pm 10\%$, due to the fact that we have considered in each case values of the vector masses $M$ that we expect could be compatible
with direct LHC constraints on the vector-like fermions. The correlation of $R_{Z\gamma}$ and 
$R_{\gamma\gamma}$ is markedly distinct in the 3 cases, but NP effects in $h 
\to Z\gamma$ at the $10\%$ level will be very challenging to probe at the LHC.

\section{Constraints from Electric and Magnetic Dipole Moments and Electro-Weak 
Precision Observables}\label{EDMMDM}

\noindent
In addition to the need for a low UV cut-off, models in which
the vector leptons share all quantum numbers
with the SM leptons induce 1-loop contributions to SM fermion EDMs and
MDMs. Measurements of these quantities
result in very constraining limits, especially EDM measurements, which already
probe electro-weak 2-loop 
contributions~\cite{Hudson:2011zz,Kara:2012ay,Griffith:2009zz,Baker:2006ts}. As 
a consequence, the mixing operators in these models must have very small
coefficients or must be
forbidden by an additional symmetry. Remarkably, in both scenarios discussed in
this work, the leading contributions to EDMs and MDMs are automatically lifted 
to the 2-loop level.

\subsection{Electric Dipole Moments}

\begin{figure}[tb]
 \begin{center}
\begin{tikzpicture}[rotate=180,baseline=0,scale=1.2]
\draw[decoration={coil,aspect=0,segment
length=2.5mm,amplitude=.7mm},thick,decorate] (-2,.8) to (-1,.8);
\draw[decoration={coil,aspect=0,segment
length=2.5mm,amplitude=.7mm},thick,decorate] (-2,-.8) to (-1,-.8);
\draw[thick] (0,0) to (-1,0.8);
\draw[thick]  (0,0) to (-1,-.80);
\draw[decoration={coil,aspect=0,segment
length=2.5mm,amplitude=.7mm},thick,dashed]  (0,0) to (1,0);
\draw[thick] (2,0) to (1,0);
\node[] at (2.2,0){$\ell$};
\node[] at (-3.2,0.8){$\ell$};
\fill[color=black!] (1,0) circle (0.3ex);
\draw[thick] (-2,.8) to (-3,.8);
\fill[color=black!] (-2,.8) circle (0.3ex);
\draw[thick]  (-2,0.8) .. controls (-1,2) and (-.2,2) .. (1,0);
\draw[thick] (-1,-0.8) to (-1,0.8);
\draw[thick] (-1,-0.8) to (-1,0.8);
\node[] at (-2.2,-1) {$\gamma$};
\fill[color=black!] (0,0) circle (0.3ex);
\fill[color=black!] (-1,-0.8) circle (0.3ex);
\fill[color=black!] (-1,0.8) circle (0.3ex);
\end{tikzpicture}
\end{center}
\caption{\small Barr Zee diagram  with the Higgs diphoton sub-diagram, 
contributing to
the EDM and MDM of SM leptons.}
\label{fig:BarrZee}
\end{figure}
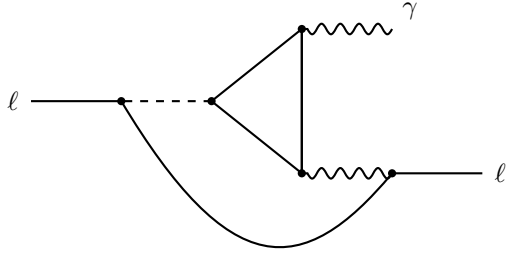

\noindent
For both $n=1$ and $n=2$, we can estimate 
contributions to the EDM of a SM fermion $f$ by considering the 2-loop Barr-Zee
type diagram in
Figure~\ref{fig:BarrZee}, which contains the $h\to \gamma\gamma$ loop as a
sub-diagram. Given that the Higgs couplings to the SM fermions are 
proportional to $m_f/v$, we obtain 
\begin{align}
\left(\frac{\Delta d_f}{e}\right)_{\mathrm{Barr-Zee}}=\frac{\alpha_e Q_f \,
Q_\chi^2}{8\pi^3} \, \frac{m_f}{v^2} ~\sum_i \frac{v}{m_i} 
\text{Im}(g_{h\chi_i\chi_i}) ~ g\left(\frac{m_i^2}{m_h^2}\right) ~.
\end{align}
The loop function $g$ can be found in Appendix~\ref{loopfuncs}.
The source of this 2-loop EDM is the same as the CP violating contribution to 
$h\to \gamma\gamma$, namely the irreducible phase in the mass matrix 
$\bm{\mathcal{M}}_E$. In the limit $m_i \gg v$ and for $M_L = M_E = M$ we can 
write
\begin{align}
 \left(\frac{\Delta d_f}{e}\right)_{\mathrm{Barr-Zee}} \simeq \frac{\alpha_e 
Q_f \,
Q_\chi^2}{8\pi^3} \, \frac{m_f}{v^2} ~ \big[ v\,\partial_v \arg (\det 
\bm{\mathcal{M}}_E) \big] ~ \frac{1}{2} \log\left(\frac{M^2}{m_h^2}\right)~,
\end{align}
thus making the correlation with the CP-odd contribution to the
$h\to \gamma\gamma$ decay rate in \eqref{Rgg}
manifest.
The explicit expression for the derivative was already given in~(\ref{CPodd}).
Note that the Barr-Zee contributions to the EDMs scale in the same way with the 
charge of the vector leptons, $Q_\chi$, as the NP amplitude in $h \to \gamma 
\gamma$ does.

\begin{figure}[tb]
\begin{center}
 \includegraphics[width=.46\textwidth]{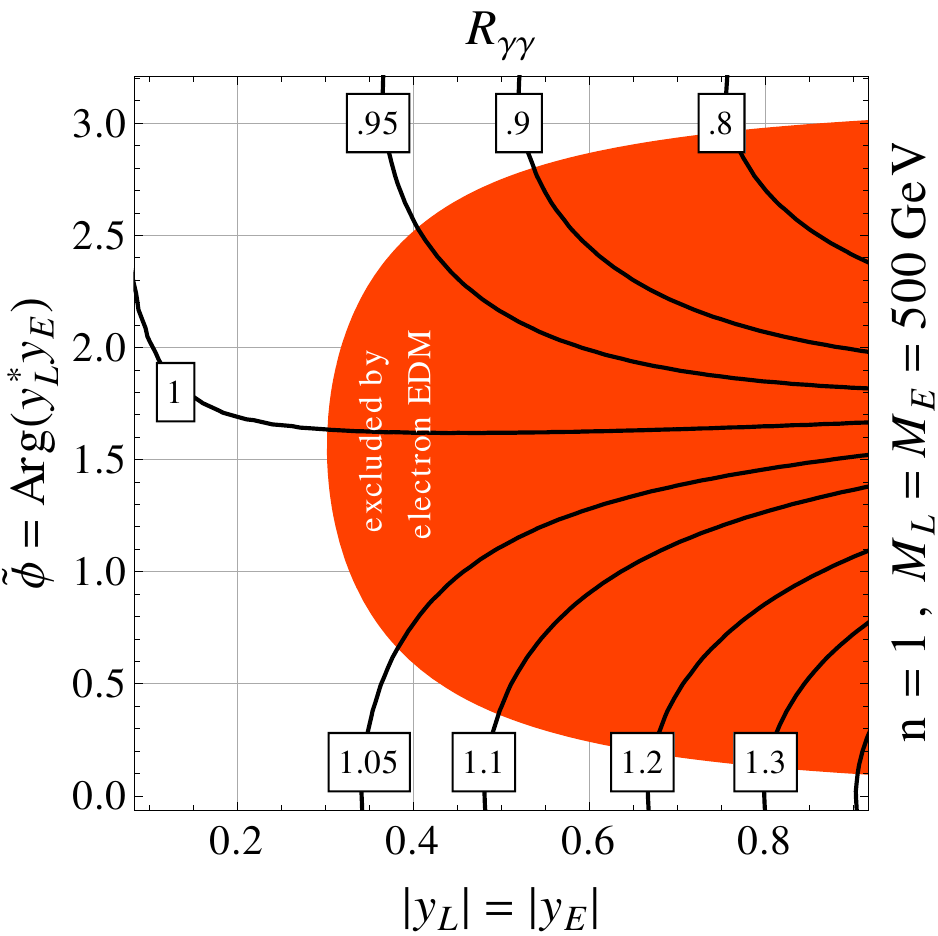} ~~~~~~
 \includegraphics[width=.46\textwidth]{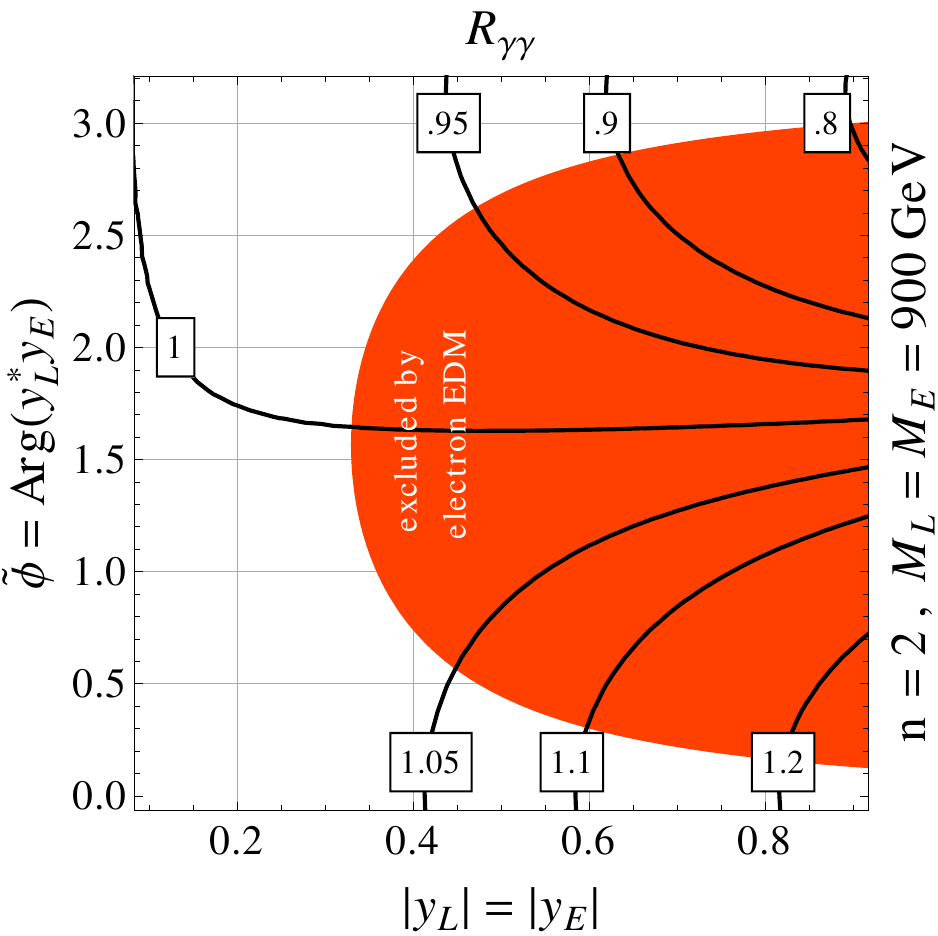}
\caption{\small
Modifications of the $h\to \gamma\gamma$ rate in the $|y| - \tilde\phi$ 
plane in the $n=1$ (left) and 
$n=2$ (right) scenarios. Vector masses are fixed to 500~GeV and 900~GeV, 
respectively. The region excluded by the electron EDM is shown in orange.}
\label{fig:EDMPlot} 
\end{center}
\end{figure}

As we will show,
bringing the 2-loop contributions in agreement with the most recent 
measurements of the electron EDM~\cite{Hudson:2011zz,Kara:2012ay},
\begin{equation}
\label{eq:de_exp}  d_{e} \leq 1.05 \times 10^{-27} ~e\,{\rm cm}~~~@~90\%~ 
\textnormal{C.L.} ~,
\end{equation}
still requires a fine-tuning of the phase of about $10\%$, in regions of 
parameter space that allow for a sizable modification of the CP conserving part 
of the $h \to 
\gamma\gamma$ amplitude, see e.g. Figure~\ref{fig:EDMPlot}.
Experimental results on EDMs of hadronic systems, e.g. the neutron EDM or 
mercury EDM~\cite{Griffith:2009zz,Baker:2006ts}, lead to constraints on quark 
EDMs that translate into comparable bounds on the model parameters, but they 
are subject to large hadronic uncertainties. 
Note, that
additional diagrams with the internal $h\gamma$ replaced by a $hZ$ can
be important for quark EDMs, but will play essentially no role for leptons 
because of the
accidentally small vector coupling of the $Z$ to SM leptons.
2-loop diagrams with 
$W^+W^-$ in the loop turn out to be small for both quarks and 
leptons, see also~\cite{McKeen:2012av}. 
Nonetheless, in our numerical analysis, we take into 
account the full set of $h\gamma$, $hZ$, and $W^+W^-$ contributions.  

\begin{figure}[tb]
\begin{center}
\includegraphics[width=0.44\textwidth]{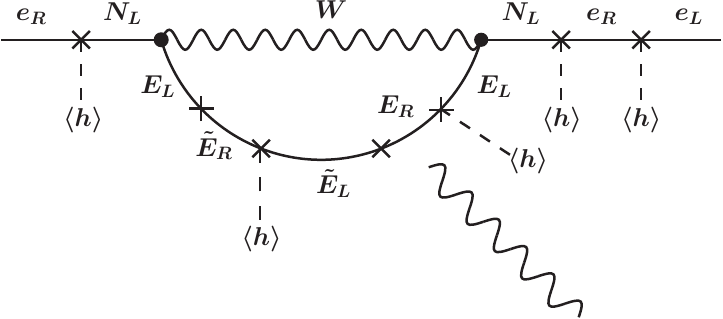} ~~~~~~~~
\includegraphics[width=0.37\textwidth]{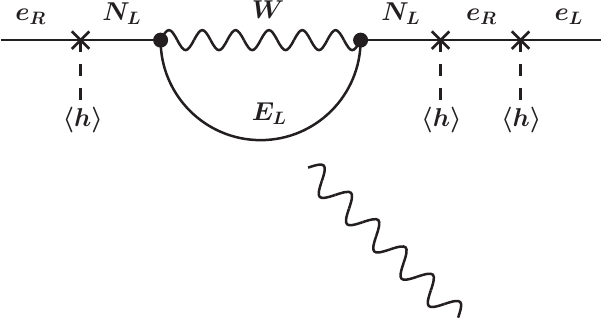}
\end{center}
\caption{\small Example 1-loop diagrams giving rise to an electron EDM (left) 
and 
MDM (right) in the $n=1$ scenario. The photon can be attached to all charged 
particles in the loops}
\label{1loop}
\end{figure}

The mixing operator in the $n=1$ scenario
also allows for an additional
1-loop contribution to the SM lepton EDMs. As the new states only mix with 
right-handed SM leptons, the physical phases in the mixing Yukawas cannot be 
accessed at the 1-loop level. The only possible 1-loop contribution is 
therefore a loop of a $W$ boson and charge two vector
leptons that is sensitive to the phase in the charge 2 mass matrix 
$\bm{\mathcal{M}}_E$. This 1-loop EDM corresponds to a dimension 10 operator, 
containing 5 Higgs fields (see diagram on the left hand side of 
Figure~\ref{1loop}). It can only compete with the 2-loop dimension 6 
contribution if $|y_{L\ell}|=\mathcal{O}(1)$ and $M_L ,M_E \simeq v$.
Given the constraints on the mixing Yukawas discussed in 
Section~\ref{constraints}, we find that the 1-loop contribution is completely 
negligible. 

Figure~\ref{fig:EDMPlot} illustrates the impact of the electron EDM bound on a 
modified $h\to\gamma\gamma$ rate. Shown are modifications of the Higgs diphoton 
rate in the $|y_L| = |y_E|$ vs $\tilde \phi = \text{Arg}(y_L^* y_E)$ plane.
The left (right) plot shows the $n=1$ ($n=2$) case with the mass of the vector 
leptons fixed to exemplary values of $M = M_L = M_E = 500$~GeV ($900$~GeV). The 
parameter space ruled out by EDMs 
generated from the 2-loop
Barr-Zee diagrams is shown in orange. Away from the limits $M_L = M_E$ and 
$|y_L| = 
|y_E|$, the results do not change qualitatively. We observe that O(1) phases 
are allowed, but 
only for small values of the couplings that cannot lead to any appreciable 
modification of the $h\to\gamma\gamma$ rate. A non-negligible modification of 
$h \to \gamma\gamma$ is only possible if the phase is at most at the 
level of $0.1$. Correspondingly, a CP-odd contribution to the $h\to \gamma\gamma$
rate at the percent level would already be in conflict with 
EDMs, barring accidental cancellations with contributions induced by additional 
CP violating sources from beyond the models considered here.
This agrees with the findings 
in~\cite{McKeen:2012av,Fan:2013qn}.
Analogously, EDM bounds also strongly restrict possible CP violating effects in 
$h \to Z \gamma$ well below the percent level. 
Possible CP violation in the experimentally most favorable $h \to ZZ$ channel 
is even further suppressed below the $10^{-4}$ level, because loop induced CP 
violating effects have to compete with the CP conserving tree level $hZZ$ 
coupling.
Since the imaginary part of the couplings is constrained to be very small,
we will only work with real $y_L$ and $y_E$ couplings for the remainder of
this paper.

\subsection{Anomalous Magnetic Moments}

\noindent
The 2-loop Barr-Zee diagrams also give contributions to anomalous magnetic 
moments of leptons in both scenarios
\begin{align}
(\Delta a_\ell)_{\mathrm{Barr-Zee}}= \frac{\alpha_e \,
Q_\chi^2}{4\pi^3} \, \frac{m_\ell^2}{v^2} ~\sum_i \frac{v}{m_i} 
\text{Re}(g_{h\chi_i\chi_i}) ~ f\left(\frac{m_i^2}{m_h^2}\right) ~.
\end{align}
with the explicit form of the 2-loop function $f$ given in 
Appendix~\ref{loopfuncs}.
In the limit $m_i \gg v$ and for $M_L = M_E = M$ we can write
\begin{align}
(\Delta a_\ell)_{\mathrm{Barr-Zee}} \simeq \frac{\alpha_e \,
Q_\chi^2}{4\pi^3} \, \frac{m_\ell^2}{v^2} ~ \big[ v\,\partial_v \log (\det
\bm{\mathcal{M}}_E^\dagger \bm{\mathcal{M}}_E) \big] ~ \frac{1}{3} 
\log\left(\frac{M^2}{m_h^2}\right)~.
\end{align}
This shows clearly the correlation of the anomalous magnetic moments with the 
CP-even contributions to the $h\to \gamma\gamma$ decay rate. The explicit 
expression for the derivative can be found in~(\ref{CPeven}).

However, given the uncertainty of the current experimental results 
and the precision of the SM predictions~\cite{Giudice:2012ms}
\begin{eqnarray}
 \Delta a_\mu &=& a_\mu^{\rm exp} - a_\mu^{\rm SM}  = (2.9 \pm 0.9) \times
10^{-9} ~, \\
 \Delta a_e &=& a_e^{\rm exp} - a_e^{\rm SM}  = (-10.5 \pm 8.1) \times 10^{-13}
~,
\end{eqnarray}
we find that the 2-loop contributions lead to effects that are one 
order of magnitude below the current sensitivities or even smaller, even for 
vector masses at 
the order of the electro-weak scale and Yukawa couplings of order 1. 

In the $n=1$ scenario, there are in addition various 1-loop contributions 
coming 
from
Higgs, $W$ and $Z$ exchange between the SM and the vector leptons. 
In contrast to the 1-loop contribution to the EDMs, the 1-loop MDMs correspond 
to dimension 8 operators (see the example diagram on the right hand side in 
Figure~\ref{1loop}).
Nevertheless, we find that only for $M_L, M_E \sim v$ and 
$y_{L\ell}=\mathcal{O}(1)$ can 1-loop MDMs reach the current sensitivities.
For all realistic choices of parameters, all contributions to the MDMs are 
negligible in our models.\footnote{Sizable contributions to the anomalous 
magnetic moments of leptons can arise in 
models where the vector-like leptons have the same quantum numbers as the SM 
leptons~\cite{Kannike:2011ng,Dermisek:2013gta}.}

\subsection{S and T Parameter} \label{sec:ST}

Additional constraints on the discussed scenarios arise from electro-weak 
precision observables, in particular the S and T parameters. The latest 
constraints on S and T read~\cite{Baak:2012kk}
\begin{equation}
 \Delta S = 0.03 \pm 0.10 ~,~~~  \Delta T = 0.05 \pm 0.12 ~,
\end{equation}
with a strong positive correlation between the two parameters of $+0.89$.
In our setups, contributions to S and T arise at 1-loop and at order
$O(y^4 v^2/M^2)$, with $M \sim M_L, M_E$ and $y \sim y_L, y_E$.

\begin{figure}[tb]
\begin{center}
 \includegraphics[width=.48\textwidth]{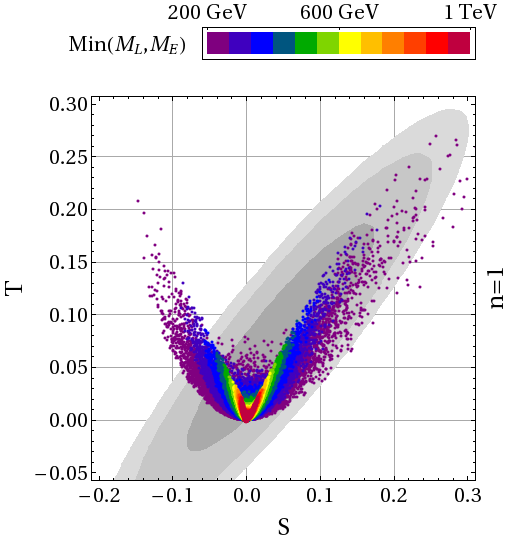} ~~~~
 \includegraphics[width=.48\textwidth]{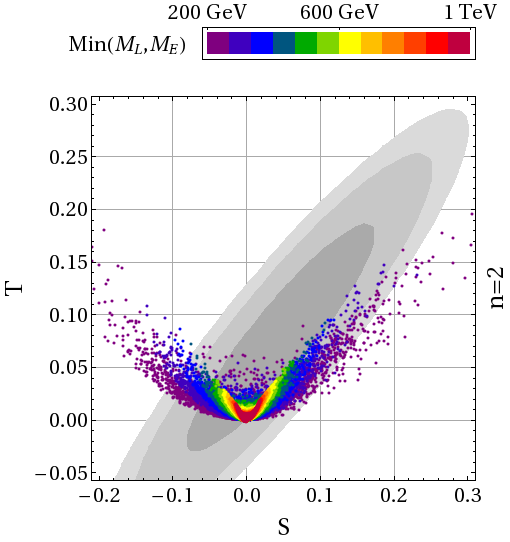}
\caption{\small
Contributions to the S and T parameters from the new vector-like leptons for 
$n=1$ (left plot) and $n=2$ (right plot).
In both plots, the Yukawa couplings are varied in the range $-1 < y_E, y_L < 
1$. The color code indicates the value of the lighter of the vector masses. The 
region allowed by the electro-weak precision fit is shown with the gray 
ellipses at the 1, 2, and 3$\sigma$ level.}
\label{fig:ST} 
\end{center}
\end{figure}

Contributions to the T parameter are independent of the hypercharge of the new 
vector-like leptons. The T parameter leads only to weak 
constraints on the masses $M_L$ and $M_E$ and the Yukawa couplings $y_E$ and 
$y_L$. Even for sizable 
Yukawas, $y_E = y_L = 1$, vector masses as low as $M_L = M_E = 300$ GeV are 
allowed~\cite{Kearney:2012zi}. 
Contributions to the S parameter do depend on the hypercharge 
assignments. We calculate corrections to the S parameter in our scenarios by 
adapting the general expressions given in~\cite{Lavoura:1992np}.
We find that despite the large hypercharges, corrections to the S parameter are 
typically also 
moderate. 

This is illustrated in Figure~\ref{fig:ST} which shows the 
contributions to the S 
and T parameters from the new vector-like leptons for $n=1$ (left plot) and 
$n=2$ (right plot). 
In both plots, the Yukawa couplings are varied independently 
in the range $-1 < y_E, y_L < 1$ 
and the vector masses are $M_L , M_E > 200$~GeV. The color code indicates the 
value of the lighter of the vector masses.  
The region allowed by the electro-weak precision fit~\cite{Baak:2012kk} is 
shown with the gray ellipses at the 1, 2, and 3$\sigma$ level.
In the regions of parameter space that are not excluded in the minimal models 
by direct searches ($M_L, M_E \gtrsim 370$~GeV for $n=1$ and $M_L, M_E \gtrsim 
850$~GeV for $n=2$, see Section~\ref{secdecay}), 
we typically have very small corrections, $\Delta T \lesssim 0.05$ and 
$\Delta S \lesssim 0.05$, well within the range allowed by the precision 
electro-weak fit.
We find that the S and T parameter can lead to non-trivial constraints 
only for very small vector masses of $\sim 200$~GeV$ - 300$~GeV.

\section{Constraints on Mixing with the Standard Model Leptons} 
\label{constraints}

\noindent
In the $n=1$ case, the mixing between the SM leptons and the new leptons is 
subject to strong indirect constraints from $Z$ pole observables and lepton 
flavor violating processes.
In this section, we discuss the most stringent constraints and their 
implications. 

\subsection{Constraints from Modified Z and Higgs Couplings}

\begin{figure}[tb] \centering
\includegraphics[width=0.45\textwidth]{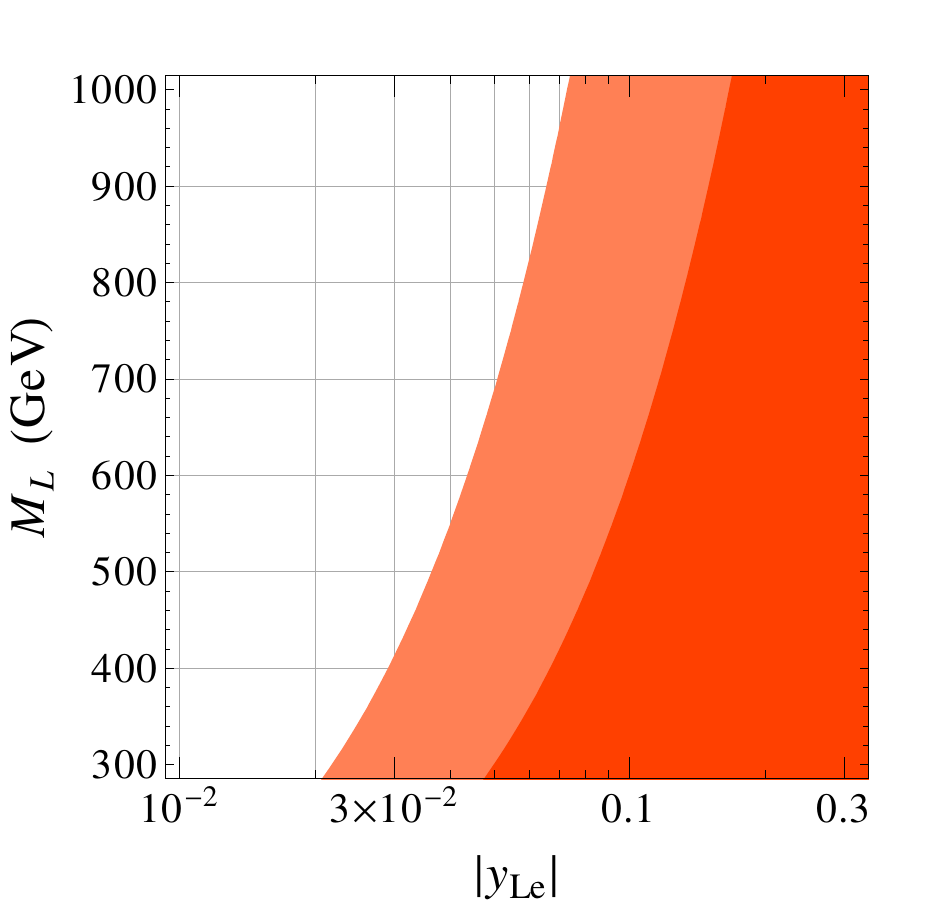} ~~~~~~
\includegraphics[width=0.45\textwidth]{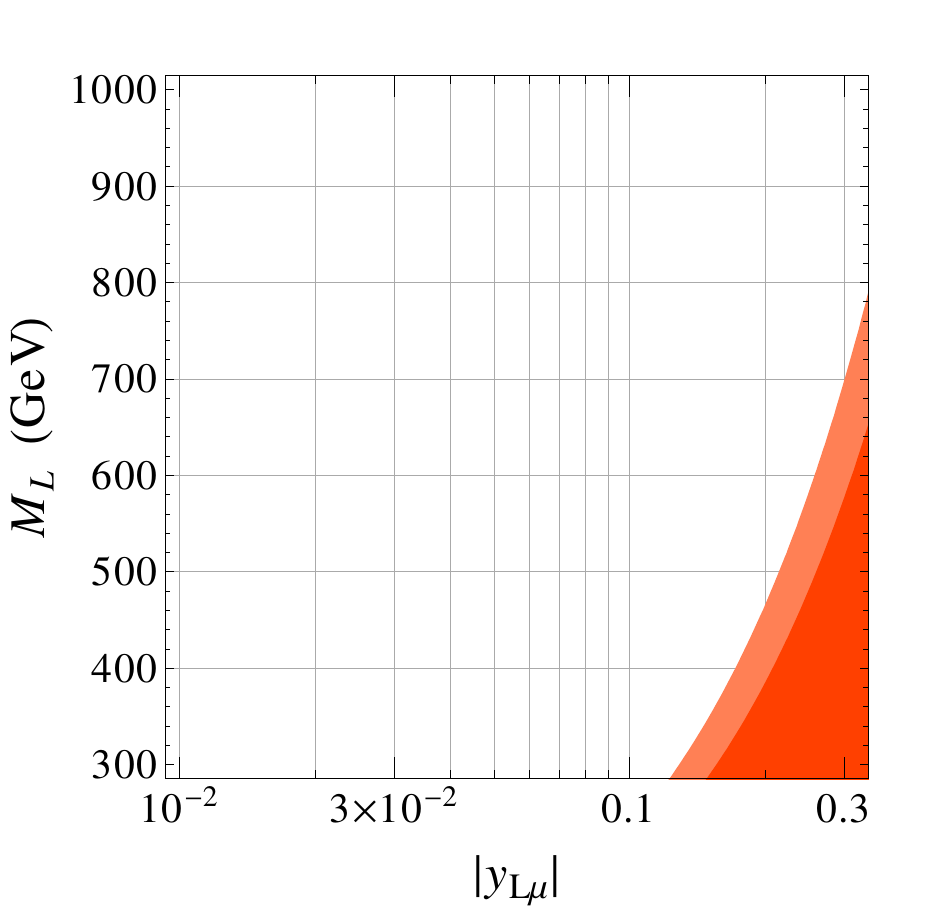} \\
\includegraphics[width=0.45\textwidth]{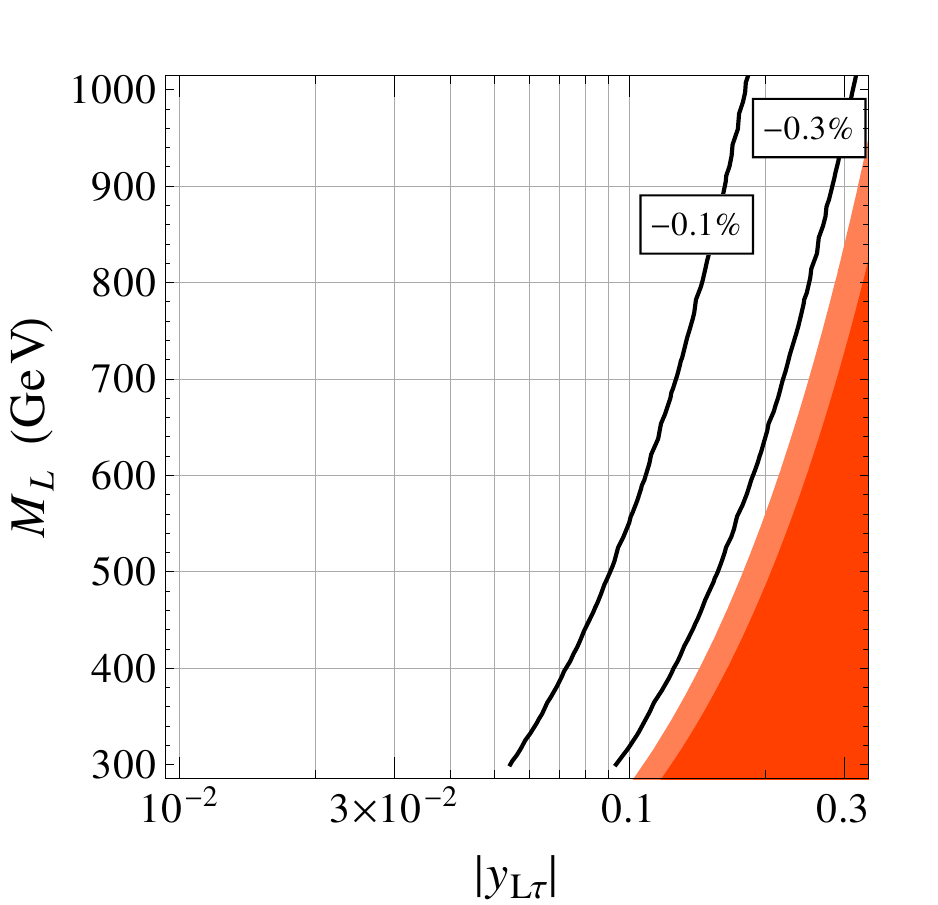} ~~~~~~
\caption{\small
Constraints from the LEP measurements of the $Z$ boson couplings in the plane 
of 
the vector lepton mass $M_L$ and the mixing Yukawa couplings $y_{L\ell}$ with 
electrons (top left), muons (top right), and taus (bottom). The light (dark) 
orange regions are excluded at the $2\sigma$ ($3\sigma$) level. In the bottom 
plot, modifications of the $h \to\tau\tau$ rate are also shown with the 
black contours.
}
\label{fig:Zpole}
\end{figure}

\noindent
The couplings of the SM leptons to the $Z$ boson have been precisely measured 
at 
LEP. As already mentioned at the end of Section~\ref{models}, the Yukawa 
couplings that mix the SM leptons with the new particles lead to corrections 
to the coupling of the $Z$ with the right-handed SM leptons.
Such corrections are constrained at the $10^{-3}$ level and 
better~\cite{ALEPH:2005ab}. Combining the experimental results with the SM 
predictions collected in~\cite{ALEPH:2005ab} we find
\begin{eqnarray}
 \delta g_{Re} &=& |y_{L e}|^2 \frac{v^2}{2M_L^2} = -0.00060 \pm 0.00034~, \\ 
 \delta g_{R\mu} &=& |y_{L \mu}|^2 \frac{v^2}{2M_L^2} = 0.0002 \pm 0.0013~, \\ 
 \delta g_{R\tau} &=& |y_{L \tau}|^2 \frac{v^2}{2M_L^2} = 0.00066 \pm 0.00064~,
\end{eqnarray}
where the $\delta g_{R\ell}$ are defined as the relative deviations of the 
coupling of the $Z$ with the right-handed SM leptons
\begin{equation}
 \mathcal{L} \supset - \frac{e}{s_W c_W}Z^\mu ~ \bar \ell ~ \gamma_\mu \Big[ 
\left( g_{L\ell}^\text{SM} + \delta g_{L \ell }\right) P_L + \left( 
g_{R\ell}^\text{SM} + \delta g_{R \ell }\right) P_R \Big] ~\ell~.
\end{equation}
The model predicts always positive corrections to the couplings $g_{R\ell}$. As 
the 
measured coupling of electrons is almost $2\sigma$ below the SM prediction, the 
derived constraints are particularly strong in the case of electrons.
The constraints are illustrated in the plots of Figure~\ref{fig:Zpole} in the 
$M_L$ - $y_{L\ell}$ planes. Dark and light orange regions are excluded at the 
3$\sigma$ and 2$\sigma$ level, respectively.

There can in principle be also corrections to the decay of the Higgs to leptons.
We find at leading order the following modification of the $h\to\tau\tau$ 
signal strength
\begin{equation}
 R_{\tau\tau} \simeq 
\frac{\Gamma(h\to\tau\tau)}{\Gamma(h\to\tau\tau)_\text{SM}}  \simeq 1- |y_{L 
\tau}|^2 \frac{v^2}{M_L^2}~.
\end{equation}
Contours of constant $R_{\tau\tau}$ are superimposed in the bottom plot of 
Figure~\ref{fig:Zpole}. Given the constraints from the $Z$ pole measurements, 
this correction is unobservably small. 
This is in contrast to the $n=0$ case where the additionally allowed mixing 
Yukawas and masses can lead to visible modifications of Higgs couplings to 
fermions~\cite{Kearney:2012zi}.

\subsection{Lepton Flavor Violation}

\noindent
Very stringent constraints on the coefficients of the mixing operators in
the $n=1$ Lagrangian also come from observables
measuring the flavor changing couplings of the $Z$. The most
severe bounds result from the tree-level induced $\mu \to e$ conversion in 
nuclei, and flavor violating $\tau$ decays, like $\tau \to 3 e$
and $\tau \to 3 \mu$. 

For the $\mu \to e$ conversion in nuclei, the
branching ratio can be written as~\cite{Kitano:2002mt}
\begin{equation}
 {\rm BR}(\mu \to e ~\text{in~N}) \times \omega_{\rm cap.}^N = 4 \Big| (2C_u + 
C_d)
V^{(p)} +(C_u + 2C_d) V^{(n)}\Big|^2 ~,
\end{equation}
in which $\omega_{\rm cap.}^N$ denotes the muon capture rate of the nucleus $N$,
and $V^{(p)}$ and $V^{(n)}$ are nucleus dependent overlap
integrals~\cite{Kitano:2002mt}.

The coefficients $C_u$ and $C_d$ are defined by the effective Hamiltonian
\begin{equation}
 \mathcal{H} = C_q (\bar e \gamma_\nu P_R \mu)(\bar q \gamma^\nu q) ~, 
\end{equation}
and are generated by off-diagonal $Z$ couplings. We find 
\begin{equation}\label{Cmueconv}
C_u = \phantom{-} y_{L\mu} y_{Le}^* \frac{1}{4 M_L^2} \left( 1 - \frac{8}{3}
s_W^2 \right) ~,~~~ 
C_d = - y_{L\mu} y_{Le}^* \frac{1}{4 M_L^2} \left( 1 - \frac{4}{3} s_W^2
\right)~.
\end{equation}
The current most stringent experimental bounds are coming from measurements
using Au and Ti atoms~\cite{Kaulard:1998rb,Bertl:2006up}
\begin{eqnarray}
 {\rm BR}(\mu \to e ~\text{in~Au}) &<& \phantom{1.}7 \times 10^{-13} ~~~@~90\%~
\textnormal{C.L.}~,\\
 {\rm BR}(\mu \to e ~\text{in~Ti}) &<& 1.7 \times 10^{-12} ~~~@~90\%~
\textnormal{C.L.}~,
\end{eqnarray}
and can be translated into bounds on the combinations of couplings, which enter
\eqref{Cmueconv}. 

\begin{figure}[tb] \centering
\includegraphics[width=0.45\textwidth]{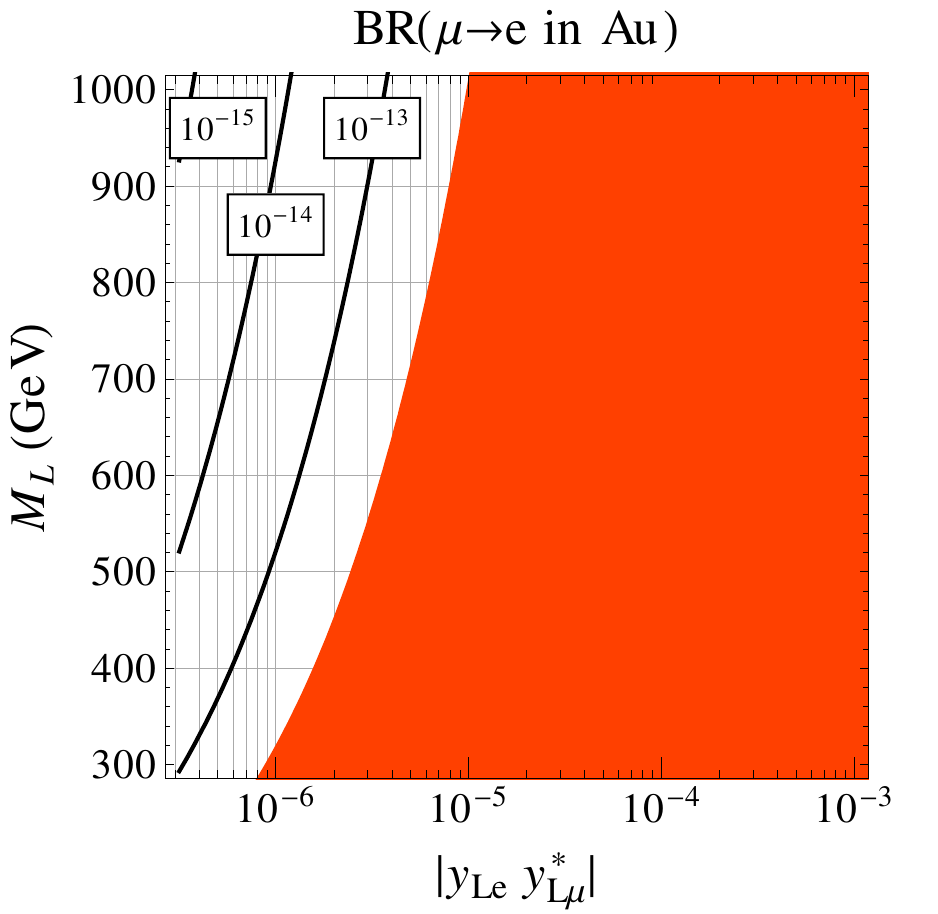} ~~~~~~
\includegraphics[width=0.45\textwidth]{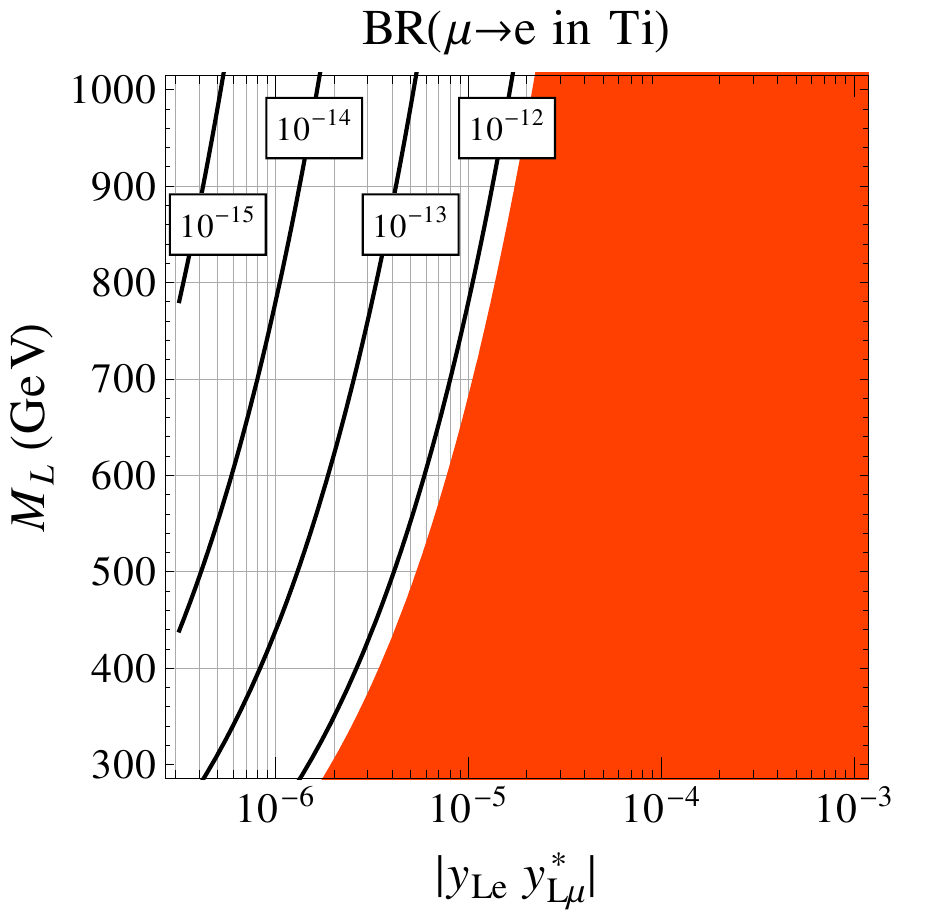}
\caption{\small
The branching ratios of $\mu \to e$ conversion in gold (left) and titanium
(right) as function of the mass of the new vector-like fermions and the relevant
combination of couplings. The orange regions are excluded by current 
constraints.
}
\label{fig:mu2e}
\end{figure}

The corresponding parameter space is shown in Figure~\ref{fig:mu2e}, with the 
excluded region shaded in orange.
Generically, for $y_{Le} \simeq y_{L\mu}$, couplings at the order of $10^{-3}$ 
are probed. However, as only the product of these two couplings is constrained, 
either of the couplings can be as large as the bound obtained from $Z$ pole 
observables in the previous section, as long as the other coupling is strongly 
suppressed.
The expected sensitivity of the Mu2e experiment to $\mu \to e$ 
conversion in Al, ${\rm BR}(\mu \to e ~\text{in~Al}) \lesssim 6 \times 
10^{-17}$~\cite{Abrams:2012er}, 
will probe large regions of the presently allowed parameter space.

\bigskip

For $\ell \to 3\ell'$ decays, the branching ratio can be written as
\begin{equation}
 \frac{{\rm BR}(\ell \to 3 \ell^\prime)}{{\rm BR}(\ell \to \ell^\prime \nu
\bar\nu)} = \frac{1}{4 G_F^2} \left( |C_{LL}|^2 +|C_{RR}|^2 +
\frac{1}{2}|C_{RL}|^2 + \frac{1}{2} |C_{LR}|^2 \right) ~,
\end{equation}
where the coefficients $C_i$ are defined by the effective Hamiltonian
\begin{eqnarray}
 \mathcal{H} &=& - C_{LL} (\bar \ell^\prime \gamma_\nu P_L \ell)(\bar 
\ell^\prime \gamma^\nu P_L \ell^\prime) -
C_{RR} (\bar \ell^\prime \gamma_\nu P_R \ell)(\bar \ell^\prime \gamma^\nu P_R 
\ell^\prime) \nonumber \\
 && - C_{LR} (\bar \ell^\prime \gamma_\nu P_L \ell)(\bar \ell^\prime \gamma^\nu 
P_R \ell^\prime) - C_{RL} (\bar
\ell^\prime \gamma_\nu P_R \ell)(\bar \ell^\prime \gamma^\nu P_L \ell^\prime) ~.
\end{eqnarray}
The dominant contribution comes again from the tree level exchange of the $Z$ 
boson
with its flavor violating coupling to right-handed leptons. We have
\begin{equation}
C_{RR} = y_{L\ell} y_{L\ell^\prime}^* \frac{1}{M_L^2} s_W^2 ~,~~~
C_{RL} = y_{L\ell} y_{L\ell^\prime}^* \frac{1}{2 M_L^2} \left(2 s_W^2 -1 
\right)~,
\end{equation}
and contributions to $C_{LL}$ and $C_{LR}$ are negligible, 
see eq.~(\ref{gZetaLetaL}). The resulting
branching ratio for $\mu\to 3 e$ gives a weaker bound than $\mu\to e$
conversion, while the $\tau \to 3e$ and $\tau \to 3\mu$ branching ratios allow
to constrain the mixing of the vector-like leptons with the $\tau$.

The current bounds on the $\tau$ branching ratios are \cite{Hayasaka:2010np}
\begin{eqnarray}
 {\rm BR}(\tau \to 3e) < 2.7 \times 10^{-8} ~~~@~90\%~ \textnormal{C.L.}~, \\
 {\rm BR}(\tau \to 3\mu) < 2.1 \times 10^{-8} ~~~@~90\%~ \textnormal{C.L.}~.
\end{eqnarray} 

\begin{figure}[tb] \centering
\includegraphics[width=0.45\textwidth]{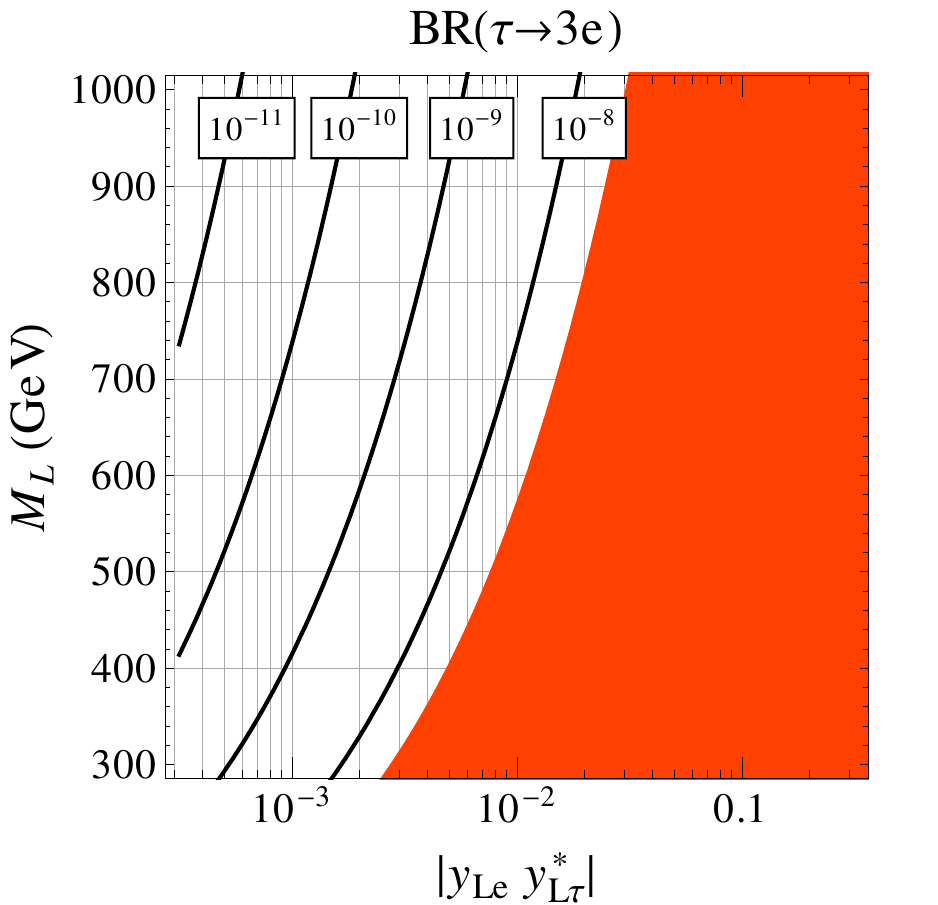} ~~~~~~
\includegraphics[width=0.45\textwidth]{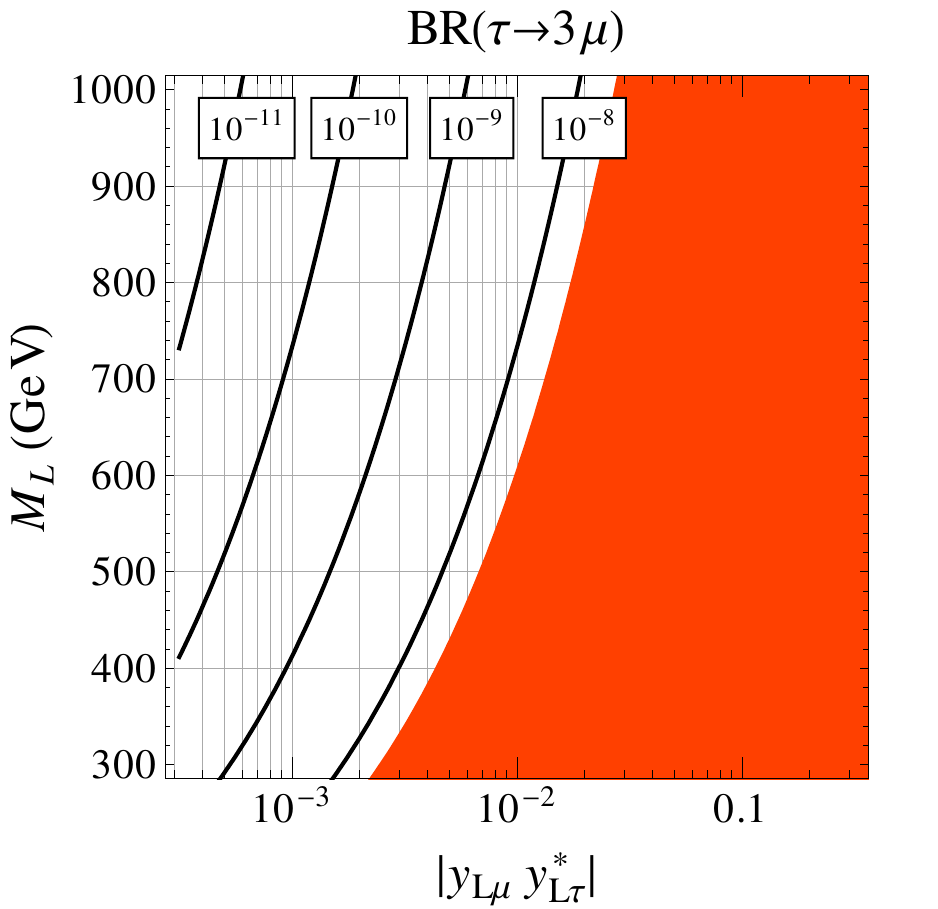}
\caption{\small
The branching ratios of $\tau \to 3e$ (left), and $\tau
\to 3\mu$ (right) as function of the mass of the new vector-like fermions and
the relevant combination of couplings. The orange regions are excluded by 
current
constraints.
}
\label{fig:mu3e}
\end{figure}

\noindent
The allowed
parameter space is shown in 
Figure~\ref{fig:mu3e}, again with the experimentally excluded region shaded in
orange. 
Other flavor violating leptonic tau decays like $\tau^+ \to e^+ \mu^+ \mu^-$,  
$\tau^+ 
\to
\mu^+ e^+ e^-$, or lepton flavor violating semi-leptonic tau decays lead to very
similar constraints. Bounds from the loop induced $\ell \to \ell' \gamma$ 
decays constrain the same combination of couplings as the observables discussed
previously, but -- due to the loop suppression -- result in much weaker constraints, 
so that we refrain from presenting a
detailed discussion of these bounds.

\begin{figure}[tb] \centering
\includegraphics[width=0.45\textwidth]{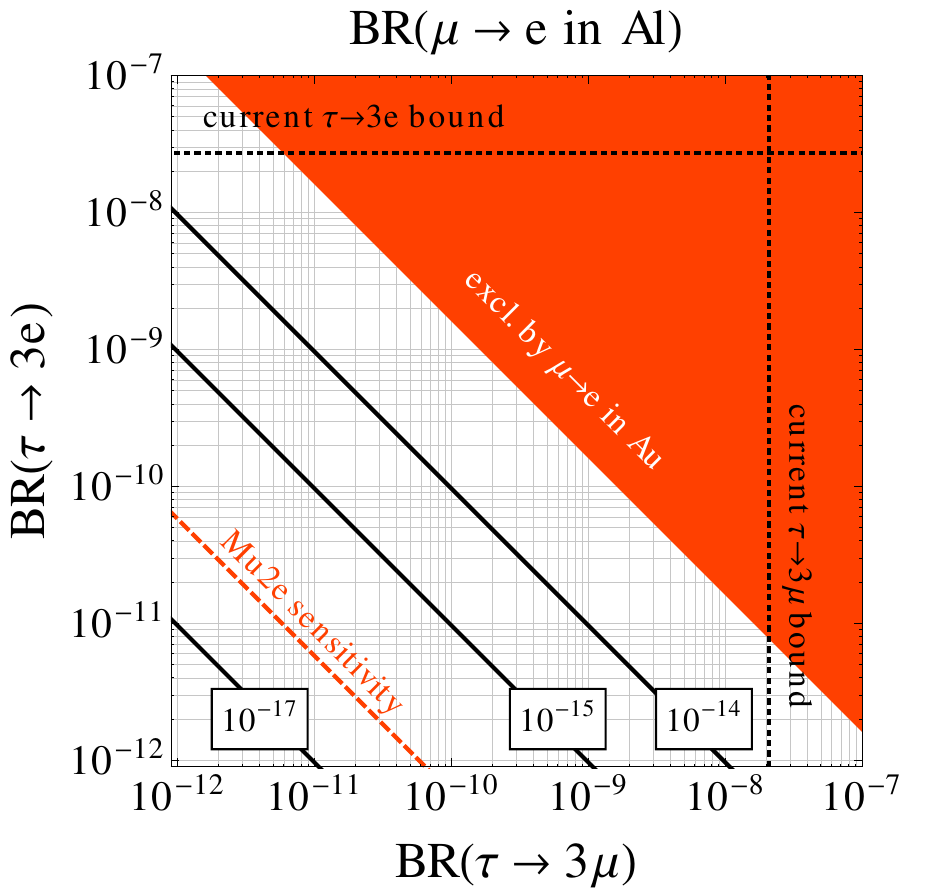} ~~~~~~
\caption{\small
Contours of constant BR$(\mu\to e ~\text{in~Al})$ in the BR$(\tau \to 3e)$ vs 
BR$(\tau \to 3\mu)$ plane. The orange region is excluded by the current 
BR$(\mu\to e ~\text{in~Au})$ constraint. The current bounds on BR$(\tau \to 
3e)$ 
and BR$(\tau \to 3\mu)$ are indicated with the black dotted lines. The 
sensitivity of the Mu2e experiment is shown with the orange dashed line.}
\label{fig:tau3e_tau3mu}
\end{figure}

Note that due to the strong constraint from $\mu \to e$ conversion either $\tau 
\to 3e$ or $\tau \to 3\mu$ can be close to the current bound, but not both 
simultaneously.
Indeed, combining the expressions for BR$(\tau \to 3e)$, BR$(\tau \to 3\mu)$, 
BR$(\mu\to e ~\text{in~Au})$, and $\delta g_{R\tau}$, we arrive at the 
following 
relation that is independent of any model parameters
\begin{equation}
 \text{BR}(\tau \to 3 e) \times \text{BR}(\tau \to 3 \mu) = \text{const.} 
\times \left( \delta g_{R\tau} \right)^2 \times \text{BR}(\mu\to e 
~\text{in~Au}) 
~.
\end{equation}
The proportionality constant is purely given by known SM parameters, and we 
find: const. $\simeq 1.2 \times 10^{-4}$.
The constraint from $\mu \to e$ conversion on possible NP effects in 
BR$(\tau \to 3e)$ and BR$(\tau \to 3\mu)$ is illustrated in 
Figure~\ref{fig:tau3e_tau3mu}. Shown in orange is the region in the BR$(\tau 
\to 
3e)$ vs. BR$(\tau \to 3\mu)$ plane that is excluded by the current bound on 
BR$(\mu\to e ~\text{in~Au})$, allowing a correction to $\delta g_{R\tau}$ that 
saturates the 
experimental $2\sigma$ upper limit. The diagonal lines indicate the values for 
$\mu \to e$ conversion in Al, in the still allowed regions. 
The expected sensitivity of the Mu2e experiment to BR$(\mu\to e ~\text{in~Al})$ 
is shown with the orange dashed line.
The current 
experimental constraints on BR$(\tau \to 3e)$ and BR$(\tau \to 3\mu)$ are shown 
with the horizontal and vertical black dotted lines.
Finding both BR$(\tau \to 3e)$ and BR$(\tau \to 3\mu)$ close to the current 
bounds 
would clearly rule out the studied framework.

In conclusion, 
observables measuring deviations of the $Z$ couplings to SM leptons lead to 
constraints on the mixing Yukawas in the $n = 1$ case.  The strongest 
bounds are summarized in Table~\ref{tab:bounds}.
The analysis in this section has to be contrasted with the results from studies
of models of new vector-like leptons, which have the exact same quantum numbers
as their SM cousins. In these models, highly non-generic CP and flavor
structures are necessary in order to satisfy the constraints discussed in this
section, see for example~\cite{Giudice:2012ms} and references therein. 

\begin{table}
\begin{center}
\renewcommand{\arraystretch}{1.6}
\begin{tabular}{r|lcr|l}
~~~~ BR$(\mu\to e ~\text{in~Au})$ ~~& ~~$\frac{v^2}{M_L^2} |y_{Le}y_{L\mu}| <
0.3 \times 10^{-6}$ ~~~&\quad\quad\quad\quad& ~~~~$\delta g_{Re}$ ~~& 
~~$\frac{v^2}{M_L^2} 
|y_{Le}|^2 < 1.6 \times 10^{-4}$ ~~~ \\
~~~~ BR$(\tau\to 3e)$ ~~& ~~$\frac{v^2}{M_L^2} |y_{Le}y_{L\tau}| <
0.9 \times 10^{-3}$ ~~~&\quad\quad\quad\quad& ~~~~$\delta g_{R\mu}$ ~~& 
~~$\frac{v^2}{M_L^2} 
|y_{L\mu}|^2 < 5.6 \times 10^{-3}$ ~~~ \\
~~~~ BR$(\tau\to 3\mu)$ ~~& ~~$\frac{v^2}{M_L^2} |y_{L\mu}y_{L\tau}| <
0.8 \times 10^{-3}$ ~~~&\quad\quad\quad\quad& ~~~~$\delta g_{R\tau}$ ~~& 
~~$\frac{v^2}{M_L^2} 
|y_{L\tau}|^2 < 3.9 \times 10^{-3}$ ~~~ \\
\end{tabular}
\end{center}
\caption{\small Summary of the strongest bounds on the mixing Yukawas from 
lepton flavor 
violating processes (left) and $Z$ pole observables (right).}
\label{tab:bounds}
\end{table}

\section{Collider Phenomenology}\label{secdecay}

\subsection{Production of the Vector-Like Leptons}

\noindent
In both scenarios, the new vector leptons will dominantly be pair produced in
Drell–Yan processes due to their large hypercharges. 
Sub-dominant channels are Higgs mediated pair production or the production of a
pair of vector-like leptons with different charges through a $W$. 

In the $n=1$ scenario, the $W$
channel does also allow for a charge 2 vector-like lepton to be produced
together with a charged SM lepton, or for the charge 1 vector-like lepton to be 
produced 
together with a SM neutrino. For $n=1$, there is also Drell-Yan production of a
charge 1 vector-like lepton together with a charged SM lepton. The single
production channels are however suppressed by the Yukawa matrix $y_{L\ell}$,
which parametrizes the mixing of the vector leptons with the SM
leptons, as well as by powers of the electro-weak scale over the vector masses.
They turn
out to be two to three orders of magnitude smaller compared to the Drell–Yan 
production at the current LHC energy
$\sqrt{s}=8$ TeV.

\subsection{Decays of the Vector-Like Leptons: $n=2$ Case}

In the minimal $n=2$ scenario, the hypercharge assignments do not allow for a 
coupling
between the new vector leptons and the SM leptons. Therefore, the lightest 
charge three
state is stable, because it cannot decay into SM fields if one considers 
the theory to be
renormalizable. 
Higher dimensional operators have to be considered within the model to 
allow for the decay of the 
lightest charge 3 state.

The lowest 
dimensional operators that can lead to a decay of the charge 
three states into SM particles are of 
dimension six, namely
\footnote{
Note, that new leptons with even larger hypercharges can only decay through
operators with dimension eight or higher.} 
\begin{align}\label{dim6op}
 \Delta \mathcal{L} \ni
&\frac{(c_R)_{ijk}}{\Lambda^2}
\,
\overline E_R \,\ell_R^i\, \ell_R^j \ell_R^k+
\frac{(c_L)_{ijk}}{\Lambda^2}
\,
\overline E_L \,\ell_R^i \,\ell_R^j \ell_L^k\,\nonumber \\[2pt]
& + \frac{(\tilde c_R)_{ijk}}{\Lambda^2}
\,
\overline{\tilde  E}_R \,\ell_R^i\, \ell_R^j \ell_L^k+
\frac{(\tilde c_L)_{ijk}}{\Lambda^2}
\,
\overline{\tilde  E}_L \,\ell_R^i \,\ell_R^j \ell_R^k\,,
\end{align}
in which a summation over lepton flavor $\ell^i,\ell^j,\ell^k=e, \mu,\tau $ is 
implicit. These operators violate SM lepton number and generically also lepton 
flavor.
 Dimension six operators that
would allow proton decay must obviously be further suppressed, which can be
achieved through a UV completion that does not couple leptons to colored 
fermions
or conserves baryon number. The most stringent bounds from
dimension six operators, which directly contribute to charged lepton flavor 
violating (LFV) processes, point to a new
physics scale of $\Lambda > 10^3$~TeV~\cite{deGouvea:2013zba}. 
If $\Lambda$ is smaller, bounds from LFV processes call for additional
flavor structure in the corresponding Wilson coefficients.
 
 Considering no additional structure in the Wilson coefficients 
$(c_L)_{ijk}\approx(c_R)_{ijk}=\mathcal{O}(1)$, the
lifetime and
decay length of the
vector leptons can be approximated by 
\begin{align}\label{lifetime}
 \tau=\frac
{1}{n}\,\frac{192\pi^3}{m_{\chi_\ell}^5}\,\Lambda^4\approx  0.2\,\text{mm}
\left(\frac{\Lambda_\mathrm{}}{10^3\, \text{TeV}
}\right)^4\,\left[\frac{800\,\text{GeV}}{m_{\chi_\ell}}\right]^5 ,
\end{align}
where a combinatorial factor $n=27$, which counts the different SM lepton
flavor variations that can appear in the operators \eqref{dim6op}, 
has been taken into account.
If the scale of new physics $\Lambda$ is sufficiently large,  $\Lambda \gg 10^3$~TeV,
the new states can behave as stable particles within the collider.
In such a case, bounds from the searches of long-lived multi-charged 
particles apply, which are approximately
$m_{\chi_\ell}\gtrsim 800$~GeV \cite{atlasnew,Chatrchyan:2013oca}. This 
translates into a bound on the vector mass of $M_L, M_E\gtrsim 850~-~970$~GeV, 
for Yukawa couplings of $y_L, y_E = 0.3-1$, leading to a stringent bound on the 
possible contribution to the $h\to \gamma\gamma$ rate once the stability of the 
vacuum is required. In particular, a 20\% enhancement can be obtained only for 
$y_L, y_E \gtrsim 0.9$, with $M_L, M_E \gtrsim 950$~GeV, and the 
Higgs quartic runs negative at scales below $\sim 10$~TeV.

The previous bounds do not apply, if the new physics scale $\Lambda$ 
is below $10^3$~TeV. In this case, generic bounds from LFV processes 
imply some mild structure in the Wilson coefficients of dimension six operators.
Depending on the flavor structure, the mass $m_{\chi_\ell}$, 
and the new physics scale, the 
lightest charge three state can decay either promptly or with a 
displaced vertex. Assuming prompt decay, from~\eqref{lifetime} one can see, 
that one order in magnitude in the new physics scale, 
$\Lambda\approx 10^2$~TeV translates to a mass limit of 
$m_{\chi_\ell}\gtrsim 125$~GeV. Even for such low masses, the UV completion 
of this model can occur at higher scales than those demanded from
vacuum stability considerations in the $n=0$ case~\cite{Joglekar:2013zya}.
Searches for many leptons~\cite{ATLAS3leptons,ATLAS4leptons,CMS1}
or leptons from displaced vertices~\cite{Chatrchyan:2012jna} can in 
principle probe larger masses for the new leptons, 
but such searches could depend strongly on 
the flavor structure. In addition, present searches require in almost 
all cases large missing momentum and are therefore not 
applicable to our case.
Since the new charge three leptons are pair produced with a 
cross-section of $\sim 100~-~0.1$~fb for masses of $m_{\chi_\ell}=300~-~1000$~GeV 
at the $\sqrt{s}=8$ TeV LHC, we expect that a dedicated analysis
utilizing existing data could already constrain the parameter space of this model. 
In particular, a search for the striking signature of six 
leptons in the final state may lead to strong constraints on 
the vector fermion mass scale.  
A detailed study of the collider phenomenology  
in these scenarios will be presented elsewhere.

\subsection{Decays of the Vector-Like Leptons: $n=1$ Case}

\begin{figure}[tb] \begin{center}
 \begin{tikzpicture}[scale=1.5,baseline=0pt]
\draw[decoration={snake,aspect=0,segment
length=2.5mm,amplitude=.7mm,},thick,decorate]  (0,0) to (1,0);
\draw[thick] (1,0) to (2,0.75);
\draw[thick] (1,0) to (2,-0.75);
\draw[decoration={snake,aspect=0,segment
length=2.5mm,amplitude=.7mm,},thick,decorate]  (2,0.75) to (2.6,1.5);
\draw[decoration={snake,aspect=0,segment
length=2.5mm,amplitude=.7mm,},thick,decorate]  (2,-0.75) to (2.6,-1.5);
\draw[thick] (2.6,1.5) to (3.2,1.75);
\draw[thick] (2.6,1.5) to (3.2,1.25);
\draw[thick] (2.6,-1.5) to (3.2,-1.75);
\draw[thick] (2.6,-1.5) to (3.2,-1.25);
\draw[thick] (2,.75) to (3.2,.75);
\draw[thick] (2,-.75) to (3.2,-.75);
\node[right] at (3.2,1.75){$\nu_\ell$};
\node[right] at (3.2,1.2){$\ell^-$};
\node[right] at (3.2,.7){$\ell^-$};
\node[right] at (3.2,-1.75){$j$};
\node[right] at (3.2,-1.2){$j$};
\node[right] at (3.2,-.7){$\ell^+$};
\node[left] at (2.4,1.3){$W^-$};
\node[left] at (2.4,-1.3){$W^+$};
\node[left] at (1.5,-.6){$\chi_{\ell}^{++}$};
\node[left] at (1.5,.6){$\chi_{\ell}^{--}$};
\node[left] at (0,0.1){$\gamma, Z$};
\end{tikzpicture}
\caption{\label{fig:n=1decay}\small Decay chain of the doubly charged vector 
lepton in the $n=1$ scenario.}
\end{center}
\end{figure}
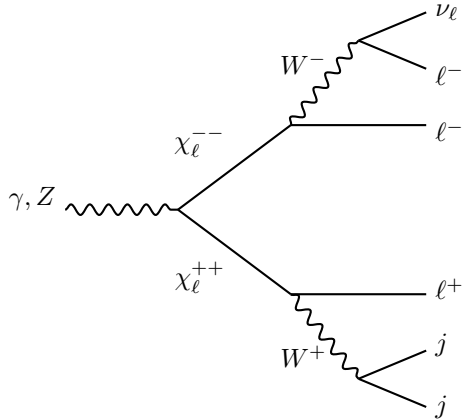

In the $n=1$ scenario, the couplings of the
vector-like leptons to SM matter allow for the direct decay of 
the lightest charge 2 state
into a SM lepton and a $W$. 
The flavor observables discussed in Section~\ref{EDMMDM} constrain combinations 
of these couplings and are collected in the left column of 
Table~\ref{tab:bounds}. The right column of 
Table~\ref{tab:bounds} shows
constraints on the 
individual Yukawa couplings from $Z$ pole observables, which result in the 
bounds 
 $y_{Le}\lesssim 0.01$ for electrons and $y_{L\mu},\, y_{L\tau}\lesssim 0.1$ 
for 
muons and taus, assuming vector masses of the order of the Higgs vev. Apart 
from 
the strong bound on the product of 
the couplings of the new charge two leptons into electrons and muons from $\mu 
\to 
e$ conversion, flavor constraints are typically also satisfied in this 
parameter region. 

The production and decay topology of the lightest charge two mass eigenstate in 
the $n=1$ scenario is 
shown in 
Figure~\ref{fig:n=1decay}. The heavier charge two and charge one mass 
eigenstates dominantly decay into the lightest charge two mass eigenstate by 
radiating of $W$s and $Z$s, because the direct decay into SM matter is 
suppressed by the Yukawa couplings $y_{L\ell}$ and powers of the electro-weak 
over the vector mass scale. Since a sizable effect in $h\to \gamma\gamma$ 
prefers mass splittings of the order of the Higgs vev, it is safe to 
concentrate on the decays of the light charge two mass eigenstate. For 
simplicity, 
and because of flavor bounds, we 
will assume two scenarios, in which either only decays to muons or only decays 
to taus are 
allowed, with the respective other two mixing Yukawas set to zero,
\begin{align}
 &\text{Only $\mu$ decays:} \hspace{1.5cm} y_{L\mu}=0.1,\qquad 
y_{Le}=y_{L\tau}=0\,,\notag\\
 &\text{Only $\tau$ decays:}  \hspace{1.5cm} y_{L\tau}=0.1,\qquad 
y_{Le}=y_{L\mu}=0\,.\notag
\end{align}
In the first scenario, in which the new resonances couple directly only to 
muons, promising 
channels to probe the model are searches for multiple light 
leptons~\cite{ATLAS3leptons,CMS1}. The second scenario (only 
coupling to taus) can be probed by 
these searches as well in the case of leptonically decaying taus, but searches 
for hadronic taus and missing energy in the 
final state are in principle also sensitive to the new 
particles~\cite{ATLAS4leptons,ATLAStaus28,ATLAStaus26,Chatrchyan:2012ira}.
We compare our signal cross section with the bounds 
from the most recent searches for three light leptons and missing energy 
\cite{ATLAS3leptons}. In the case of hadronically decaying taus, we compute the 
bounds from searches for at least two hadronic 
taus of opposite sign and missing energy \cite{ATLAStaus28}, same-sign 
dileptonic final states with at least one tau in the final state 
\cite{Chatrchyan:2012ira} as well as searches for two hadronic taus with no 
further requirement on the charges \cite{ATLAStaus26}. At last we consider an 
ATLAS search for at least one hadronic tau and three light 
leptons~\cite{ATLAS4leptons}.
 
In order to study the signal 
cross section, we implemented our model in 
\textsc{FeynRules}~\cite{Christensen:2008py} and generated events using
\textsc{MadGraph 5}~\cite{Alwall:2011uj}. In the case of tau decays, 
\textsc{PYTHIA-PGS} was used for hadronization and detector simulation 
\cite{Sjostrand:2006za}. All cuts have been applied after the detector 
simulation. For the model 
parameters, $M=M_L=M_E$ and $y=y_L=y_E\in \mathbb{R}$ has been assumed. 
Fiducial tau efficiency tables for the ATLAS detector are publicly available, 
and we find that for the hard taus required in the searches considered here, 
the \textsc{PGS} simulation yields efficiencies roughly within $10\%$ of the 
 numbers listed in Table V in \cite{Aad:2012xsa}. Since we are only interested 
in 
an estimate on the bound on the masses of the new resonances, we will not 
correct 
for these differences here.  

\subsubsection{Light Leptons in the Final State}
\begin{table}
 \begin{tabular}{cc}
  Tri-boson&$0.8\pm0.8$\\
 $ZZ$&$0.25\pm0.17$\\
 $t\bar t\, V$&$0.21^{+0.30}_{-0.21}$\\
 $WZ$&$2.1\pm 1.6$\\
 SM reducible& $1.0\pm0.4$\\\hline
  $\mathbf{\sum}$ \textbf{SM} &$\mathbf {4.4\pm 1.8}$\\
 \textbf{ Data} &$\mathbf{5}$\\\hline
 $N_\mathrm{signal}$ excl. (obs) & $6.8$ 
\end{tabular}
\qquad\qquad\qquad\qquad
\begin{tabular}{c c c}
 $m_{\chi_\ell}$ [GeV] & $N^\mu_\mathrm{signal}$ exp.& $N^\tau_\mathrm{signal}$ 
exp.\\\hline
 $100$&$52.3$& $5.6$\\
 $200$&$82.5$& $15.6$\\
 $300$&$32.6$&$8.1$\\
 $400$&$12.1$&$3.6$\\
 $500$&$4.6$&$1.5$ 
\end{tabular}
\caption{\label{table3l}
Left: expected number of background events, observed number of events in the 
data and observed limit on the number of signal 
events ($95\%$ CL) for the relevant signal region, taken 
from the ATLAS search for three final state leptons 
with $20.7$ fb$^{-1}$~\cite{ATLAS3leptons} (statistical and systematic 
errors are combined). Right: expected number of signal events 
after all cuts (see text for details) for the $n=1$ scenario with 
$y_{L\mu}=0.1,\, y_{L\tau}= y_{L e}= 
0$ denoted by $N_\mathrm{signal}^\mu$ as well as $y_{L\tau}=0.1,\, y_{L\mu}= 
y_{L e}= 
0$, denoted by $N_\mathrm{signal}^\tau$, for different masses of the lightest 
mass eigenstate.  
}
\end{table}

The $n=1$ scenario with couplings to 
light leptons
leads to the same signature expected from the electro-weak production of 
charginos and neutralinos, which subsequently 
decay into three light leptons, neutrinos and the lightest neutralino (LSP). 
In the model presented here, there is no massive neutral final 
state, so that the strong exclusion bounds for a mass-less LSP apply. If the 
new resonances are assumed to couple only to taus, this final state will also 
be a promising channel in the case that the taus decay leptonically.  

Several cuts have been imposed in  \cite{ATLAS3leptons}, in order to reduce the 
background. Exactly three light leptons are required. At least one pair of 
which must be of the same flavor with opposite charges (SFOS). In addition, the 
SFOS pair 
 with an invariant mass closest to the $Z$ mass 
must have an invariant mass of $m_\mathrm{SFOS}< 81.2$ GeV or $m_\mathrm{SFOS}> 
101.2$ GeV (``$Z$-veto'') in order to suppress $ZZ$ background. The missing 
transverse energy is required to be 
more than $E_T^\mathrm{miss}>75$ GeV. An additional 
requirement on the transverse mass of the third lepton  $m_T=\sqrt{2\, 
E_T^\mathrm{miss}\, p_T^\ell\,(1-\cos\Delta\phi_{\ell,E_T^\mathrm{miss}})}$ 
(the one which is not 
part of the $Z$ pair) of $m_T > 110 $ GeV is enforced 
in order to
suppress background from $WZ$ events. The third lepton is also required to have 
a 
transverse momentum of $p_T^\ell > 30$ GeV. 

The background predictions after cuts, number of observed 
events and upper limit on contributions from new physics \cite{ATLAS3leptons} 
are compiled on the 
left-hand side of Table~\ref{table3l}. 
We cross-checked our simulation by reproducing
the irreducible $ZZ$ background within its errors. 
The simulated number of signal events 
after cuts, depending on 
the mass of the lightest mass eigenstate of the model discussed here, is shown 
on the right-hand side of Table~\ref{table3l}. Since in our model a third 
lepton will always be the product of a $W$ decay, in the low mass region 
$m_{\chi_\ell}< 200$ GeV, the $E_T^\mathrm{miss}$ cut is the 
most efficient cut on our signal, while the requirement on $m_\mathrm{SFOS}$ 
represents the strongest cut for higher masses. 
In the case of only direct couplings to muons, 
the limits on the mass of the lightest charge 2 mass eigenstate 
are roughly $m_{\chi_\ell}\gtrsim 
460$~GeV. The scenario in which only direct couplings to taus are assumed 
allows for the weaker bound 
$m_{\chi_\ell}\gtrsim 320$~GeV.

\subsubsection{Hadronic Taus in the Final State}

In addition to the bound derived from the decay into light leptons, we 
considered three different searches for hadronic taus in the final state in 
order to further constrain the $n=1$ scenario 
in which the doubly charged leptons only decay into taus and 
$W$s. We studied searches for opposite sign hadronic taus ($+~ 
p_T^\mathrm{miss}$)~\cite{ATLAStaus28} or one hadronic tau together with a 
same-sign lepton ($+$ jets and $E_T^\mathrm{miss}$) ~\cite{Chatrchyan:2012ira} 
in the final state, as well as searches for hadronic tau pairs, jets and large 
$E_T^\mathrm{miss}$~\cite{ATLAStaus26}. Finally, we consider an ATLAS search 
for three light leptons and one or more hadronic taus in the final 
state~\cite{ATLAS4leptons}. In all cases, we reproduced the 
diboson background and find reasonable agreement with the 
simulations done in the experimental studies.

In the ATLAS search 
for opposite sign hadronic taus and missing momentum \cite{ATLAStaus28}, at 
least one 
opposite sign tau pair and no additional light leptons are required with 
transverse momenta $p_T^{\tau_1} > 40$ 
GeV and $p_T^{\tau_2} > 25$ GeV. A $Z$-veto on the invariant mass of each 
opposite sign tau pair is enforced, $m_{\tau\tau}> 91$ GeV or $m_{\tau\tau}< 
71$ GeV, as well as a veto on all events with a $b$-jet in order to reject 
background events from top quarks. In addition, the missing transverse 
energy needs to be larger than $E_T^\mathrm{miss}> 40$ GeV and the largest 
value of 
$M_{T2}^2=\min_{\slashed{p}_T=\slashed{p}_1+\slashed{p}_2}\left[\max\{m_T^2(p_T^
{
\tau_1},\slashed{p}_1),m_T^2(p^{\tau_2}_T,\slashed{p}_2)\}\right]$ computed 
among all opposite sign tau pairs needs to be larger 
than $M_{T2}> 100$ GeV in order to suppress $W^+W^-$ background.
This $M_{T2}$-cut turns out to 
reduce our signal the most. After all cuts, our signal cross section is below 
the experimental bound for the whole considered mass range 
($m_{\chi_\ell}=100 \dots 500$~GeV).

The CMS search~\cite{Chatrchyan:2012ira} for two same sign leptons requires two 
jets and missing transverse energy. In order to reduce the high 
trigger rates, a significant bound on $E_T^\mathrm{miss} > 120$~GeV and the 
scalar sum of jet transverse momenta 
$H_T=\sum p_T^\mathrm{jets} > 450$~GeV has been set in the relevant search 
region.
Transverse momentum cuts for all three lepton 
flavors, $p_T^e>10$~GeV, $p_T^\mu>5$~GeV and $p_T^\tau>15$~GeV,
and on each jet, $p_T^\mathrm{jet}> 40$~GeV have been applied and a dilepton 
invariant mass of $m_{\ell\ell}>8$~GeV has been required in order to suppress 
low-mass dilepton background. Events in which a 
third opposite-sign lepton is present and the invariant mass of any two 
opposite sign leptons lies within $\pm 15$~GeV of the $Z$ mass are vetoed. 
Our model signal cross section is strongly reduced by the stringent cuts 
on jet $p_T$s and missing energy, which can only result 
from the decay of one of the $Ws$ in our model, while the other $W$ needs to 
provide the same sign lepton. As a result, the signal cross section stays below 
the experimental limit 
throughout the considered mass range.

A search for at least one hadronic tau pair in the final state and no further 
requirements on the charges has been done at ATLAS~\cite{ATLAStaus26}. 
The aim of this search is primarily to constrain the gluino mass from decays 
through stau intermediate states.
In the search region we considered, light leptons are vetoed, but at least two 
taus with $p_T^\tau > 20$~GeV are required. A cut on the sum of the transverse 
masses  $m_T^{\tau_i}=\sqrt{2\, 
E_T^\mathrm{miss}\, 
p_T^{\tau_i}\,(1-\cos\Delta\phi_{\tau_i,E_T^\mathrm{miss}})}$ is set,
$m_T^{\tau_1}+m_T^{\tau_2}\geq 150$~GeV, in order to suppress $Z+$jets 
events, and large $E^T_\mathrm{miss}> 150$~GeV is required. At 
least two jets have to be present, with  $p_T^{j_1}>130$~GeV and 
$p_T^{j_2}>30$~GeV, as well as a large scalar sum of the transverse momenta 
of these jets and taus 
$H_T=\sum p_T^\mathrm{j_i}+\sum p_T^\mathrm{\tau_i}>900$~GeV. This $H_T$ cut, 
the 
requirement for large missing energy as well as the cut on the sum of the 
transverse tau 
masses strongly reduce our signal cross section, so that the experimental 
bounds 
do not lead to constraints throughout the scanned mass range.

\begin{table}
 \begin{tabular}{cc}
  $ZZ$&$0.19\pm0.05$\\
 $ZWW$&$0.05\pm0.05$\\
 $t\bar t\, Z$&$0.16\pm 0.12$\\
Higgs&$2.3\pm 0.06$\\
 SM reducible& $1.4\pm1.3$\\\hline
  $\mathbf{\sum}$ \textbf{SM} &$\mathbf {2.0\pm 1.3}$\\
 \textbf{ Data} &$\mathbf{4}$\\\hline
 $N_\mathrm{signal}$ excl. (obs) & $7.5$ 
\end{tabular}
\qquad\qquad\qquad\qquad
\begin{tabular}{c c }
 $m_{\chi_\ell}$ [GeV] &  $N^\tau_\mathrm{signal}$ 
exp.\\\hline
 $100$&$9.3$ \\
 $200$&$23.1$ \\
 $300$&$8.6$\\
 $400$&$3.4$\\
 $500$&$1.1$
\end{tabular}
\caption{\label{table4l}
Left: expected number of background events, observed number of events in the 
data and observed limit on the number of signal 
events ($95\%$ CL) for the relevant signal region, taken 
from the ATLAS search for three light leptons and at least one $\tau$ in the 
final state  
with $20.7$ fb$^{-1}$~\cite{ATLAS4leptons} (statistical and systematic 
errors are combined). Right: expected number of signal events 
after all cuts (see text for details) for the $n=1$ scenario with 
 $y_{L\tau}=0.1,\, y_{L\mu}= 
y_{L e}= 
0$, denoted by $N_\mathrm{signal}^\tau$, for different masses of the lightest 
mass eigenstate.  
}
\end{table}

Finally, we consider a search from ATLAS for three light leptons and (at 
least) one hadronic tau in the final state~\cite{ATLAS4leptons}. In the 
scenario discussed here, this final state requires both $Ws$ to decay 
leptonically as well as one leptonic tau and one hadronic tau. In the 
considered search region an ``extended $Z$-veto'' has been employed, which 
means, that events with pairs, triplets or quadruplets of light leptons with an 
invariant mass within $10$ GeV of $M_Z=91.2$ GeV are vetoed. In addition, 
selected events are required to either have missing energy of 
$E^T_\mathrm{miss}> 100$~GeV or an effective mass 
$m_\mathrm{eff}=E^T_\mathrm{miss}+\sum p_T^\mathrm{e_i}+\sum 
p_T^\mathrm{\mu_i}+\sum p_T^\mathrm{\tau_i}+\sum p_T^\mathrm{j_i}> 400$ GeV. 
These cuts are chosen in order to reduce the dominant backgrounds from $ZZ$, 
$WZ$ and $Z+$jets. The signal cross section is reduced by the requirement to 
have exactly three light leptons and one hadronic tau, but not very sensitive 
on the additional cuts.

The background predictions after cuts, number of observed 
events and upper limit on contributions from new physics \cite{ATLAS4leptons} 
are compiled on the 
left-hand side of Table~\ref{table4l}.  
On the right-hand side, the simulated number of signal events 
after cuts, depending on 
the mass of the lightest mass eigenstate of the model discussed here is shown. 
We estimate a bound on the mass of the lightest mass eigenstate of  
$m_{\chi_\ell}\gtrsim 310$~GeV, which is very close to the value estimated 
based on the light lepton searches.

\bigskip
Thus, after considering both scenarios -- decays of the new vector leptons only 
into one 
light lepton flavor or decays only into taus -- we find 
bounds on the mass of the lightest mass eigenstate from the search for multiple 
light leptons in the final state, while searches for hadronic taus do not 
lead to further constraints on the mass of the resonances in the second 
scenario because of various tight cuts in the 
considered searches. For the scenario with direct couplings to muons we find 
an approximate mass bound of $m_{\chi_{\ell}}\gtrsim 460$~GeV, 
while in the scenario in which only coupling to taus are assumed this mass 
is constrained to $m_{\chi_{\ell}}\gtrsim 320$~GeV. This 
translates into bounds on the vector mass of 
\begin{align} \label{bound1}
m_{\chi_{\ell}}\gtrsim 460\, \mathrm{GeV}\quad\Rightarrow\quad M_L, M_E \gtrsim 
510 - 630\,
\mathrm{GeV}\,,\quad \text{for}\quad y_L, y_E=0.3-1\,,\\ \label{bound2}
m_{\chi_{\ell}}\gtrsim 320\, \mathrm{GeV}\quad\Rightarrow\quad M_L, M_E \gtrsim 
370 - 490\,
\mathrm{GeV}\,,\quad \text{for}\quad y_L, y_E=0.3-1\,.
\end{align}
In the tau case, these bounds imply that a 30\% enhancement of the 
Higgs di-photon rate
is possible with $y_E,y_L \gtrsim 0.7$, $M_L, M_E \gtrsim 440$~GeV and 
the Higgs quartic runs negative at a scale $\Lambda \lesssim 100$~TeV.
In the muon case instead, a 30\% enhancement requires $y_E,y_L \gtrsim 1$, 
with $M_L, M_E \gtrsim 630$~GeV
and the Higgs quartic runs negative already at a scale of a few TeV.

Finally, we want to mention, that a dedicated search based on existing data, 
looking for two hadronic $W$s and two light leptons in the final state might 
lead to stronger bounds on the parameter space. 

\subsubsection{Extended Scenario}

The bounds in~(\ref{bound1}) and~(\ref{bound2}) can be 
relaxed by extending the model, such that the charge two leptons predominantly 
decay into a stable, 
neutral state with a mass close to the lightest charge two state.
This can be arranged e.g. by adding to the model a SM singlet fermion $\chi_0$ 
and coupling it to the hypercharge~2 singlet $\tilde E_L$ and right-handed SM 
leptons $\ell_R^i = e_R, \mu_R, \tau_R$ via a dimension six operator
\begin{equation} \label{eq:op}
 \mathcal{L} \supset \frac{c_{ij}}{\Lambda^2} (\bar \chi_0 \ell_R^i) 
(\bar{\tilde E}_L \ell_R^j) ~.
\end{equation}
In order for the new decay mode of $\tilde{E}_L$ to dominate over the 
decay into a $W$ boson and SM lepton, the UV scale $\Lambda$ where this 
operator 
is generated has to be sufficiently small, parametrically of order $\Lambda^4 
\lesssim M^6/(v^2 |y_{L\ell}|^2)$. For mixing Yukawas close to the bounds in 
Table~\ref{tab:bounds}, this corresponds to scales around 10~TeV. The scale can 
be much higher if the mixing Yukawas are smaller. Note that the operator 
in~(\ref{eq:op}) violates SM lepton number and generically also lepton flavor. 
As already mentioned in the $n=2$ section, bounds on
LFV dimension six operators, point to a new
physics scale of $\Lambda > 10^3$ TeV
\cite{deGouvea:2013zba}. If the operators in~(\ref{eq:op}) arise at a much 
lower scale, the non observation of charged LFV calls for additional flavor 
structure in the corresponding Wilson coefficients.

If the spectrum is sufficiently compressed, searching for the lighter 
charge~2 state is very challenging and the constraints considered previously
can be completely avoided. 
Searches for the heavier charge~2 state and the charge~1 state might 
be more promising in that case. 
Due to their larger masses and correspondingly 
smaller production cross sections, however, we expect that
constraints from current searches are still much 
weaker, compared to the minimal model. A detailed study of the collider 
phenomenology of the extended $n=1$ scenario is left for future work.
See also~\cite{Alloul:2013raa} for a very recent collider study of a related 
framework.

\section{Conclusions}

\noindent
In this work, we have analyzed an extension of the SM by one generation of new
vector-like leptons with exotic hypercharges. 
We considered two models: One, in which the
hypercharges of the new electro-weak doublets and singlets are given by 
$Y=Y_\mathrm{SM}-n$, with $n=1$ and $Y_\mathrm{SM}$ denoting the hypercharge 
of the SM leptons, and one in which $n=2$. In both
scenarios, sizable enhancements and suppressions of the $h\to \gamma\gamma$ 
decay rate are possible, depending on the relative sign of the Yukawa couplings 
of the new leptons.
We did not consider scenarios with even higher hypercharges, $n \geq 3$, as in 
such cases
the hypercharge gauge coupling will develop a Landau pole at very low scales
$\lesssim 10^4$~TeV.

The bound on the scale at which one expects a UV completion due to vacuum
instability considerations can in principle be relaxed considerably compared to 
a model in
which the quantum numbers of the new leptons are copies of the SM leptons
($n=0$). In the minimal $n=1$ and $n=2$ models, however, constraints from 
direct searches put strong bounds on the 
masses of the new leptons, which in turn constrain the possible modifications to $h\to \gamma\gamma$  as a function of their Yukawa couplings.
In the $n=2$ scenario, 
an enhancement of $R_{\gamma\gamma}\simeq 1.2$, for 
example,
can be accommodated for
 new  leptons of mass of order 800 GeV for a vector mass of $M=950$~GeV and a UV scale of
$\Lambda_\mathrm{UV}\lesssim 10$~TeV. For $n=1$,
a 30\% enhancement of $R_{\gamma\gamma}$ can occur  for new  leptons of mass of order 320 GeV for a vector mass of  $M=440$~GeV and a UV scale of 
$\Lambda_\mathrm{UV}\lesssim 100$~TeV. 
In  the widely discussed $n=0$ scenario, instead, a 30\% enhancement of $R_{\gamma\gamma}$ for  a similar UV scale, $\Lambda_\mathrm{UV}\lesssim 100$~TeV,  would require
  new vector leptons as low as the Higgs mass with $M \simeq 250$~GeV, while  slightly larger value of the new lepton masses, of order of the top quark mass, with vector masses $M \simeq 350$~GeV, would call for new bosonic
degrees of freedom at the TeV scale.

Due to the exotic hypercharge assignments in the $n=1$ and $n=2$ cases, 
possible modifications of the $h \to Z\gamma$ rate can be larger 
compared to the $n=0$ case. Still, we find that corrections to the $h \to 
Z\gamma$ rate typically do not exceed $10\%$. Precision measurements of 
the $h \to Z\gamma$ and $h\to\gamma\gamma$ rates can in principle distinguish 
between the considered cases, but it will be very challenging to achieve the 
required precision at the LHC. 

We further discussed 
the new physics contributions to electric and magnetic dipole moments. 
The non-standard hypercharge assignments
strongly restrict the possible mixing operators with SM leptons, so that 
the leading contribution to the electron and quark EDMs only appear at 2-loop 
for both the $n=1$ and $n=2$ scenario.
The corresponding Barr-Zee diagrams contain the $h\to \gamma\gamma$ loop as
a sub-diagram, and a modification of $R_{\gamma\gamma}$ is therefore 
correlated with
a 2-loop contribution to EDMs and MDMs. This correlation allows in principle 
to constrain 
the imaginary part (EDMs) and the real part (MDMs) of the Yukawa couplings 
between the
new leptons and the Higgs using the very precise measurements of these
observables. We find, that the single new phase entering the contributions to 
EDMs in our setups has to be
below the order of 10\% in regions of parameter space with visible 
modifications of the $h \to \gamma\gamma$ rate. It should be stressed that 
-- barring cancellations --
this implies that CP violation in $h \to \gamma\gamma$ and $h \to Z\gamma$ is 
constrained to be well below the 1\% level. Similarly, EDMs constrain the 
possible signature 
of a pseudoscalar component of the
Higgs detectable in the $h \to ZZ$ channel to be well below the $10^{-4}$ 
level. Bounds
from MDMs turn out to be much weaker and do not constrain the interesting
parameter space, given the current precision of the experimental results and 
the SM predictions. 

In the $n=1$ scenario, the allowed mixing of the vector-like leptons with SM 
leptons leads to modifications of the couplings of the $Z$ boson to SM leptons. 
The mixing is therefore constrained both by $Z$ pole observables and by flavor 
observables like $\mu \to e$ conversion in nuclei and $\ell \to
3\ell'$ decays. We have computed the most important of these
bounds and find that the mixing 
of the vector-like leptons with SM leptons has to be generically small.
In particular, the current bounds on $\mu \to e$ conversion strongly 
constrain a simultaneous 
mixing with electrons and muons. The planned Mu2e experiment can improve this 
bound by orders of magnitude.

Finally, we discussed the collider signals of the two models, that we already 
utilized to evaluate the possible modifications to $h \to \gamma\gamma$ and $h \to Z\gamma$. The dominant 
production cross section is the pair production of the lightest charge two 
(three) state in the case of $n=1$ ($n=2$). In the $n=2$
scenario, the decay of the charge three state can only be mediated 
through higher dimensional
operators. Possible dimension six operators violate the SM lepton number and 
generically also 
violate lepton flavor. If we assume that these operators are suppressed by a 
scale 
sufficiently high such that the new leptons are metastable at collider scales,
bounds from searches for stable charged particles apply and the lightest mass 
eigenstate has to be heavier than about $m_{\chi_\ell}\gtrsim 800$ GeV.
It is possible that this bound could be softened in a modified scenario, where 
the 
higher dimensional operators arise at scales low enough, such that the lightest 
charge three states decay promptly inside the detector. Further studies would 
be 
necessary to explore this scenario.

For $n=1$, the pair produced charge two leptons can lead to final 
states with two or more 
leptons and missing energy. We studied the leading production of the lightest 
charge two mass eigenstate and the subsequent 
decay into $W$s and SM leptons. We assume only couplings to one lepton family 
in order to avoid bounds from lepton flavor violation. If the new vector 
leptons couple only to muons we find that searches for multiple 
light leptons and missing energy in the final state constrain the mass of this 
lightest state to be heavier than about
$m_{\chi_\ell}\gtrsim 460$~GeV. The same analysis for a scenario in which only 
couplings to taus are assumed yields a weaker bound of $m_{\chi_\ell}\gtrsim 
320$~GeV. A search for one hadronic tau and three light leptons leads to very 
similar bounds for the scenario in which only tau couplings are present.

We also studied the possibility of hadronic tau pairs in the final state and 
conclude that the present searches are not sensitive to our model.
 Future 
multi-lepton searches at LHC~13 as well as dedicated searches for
2 leptons and 2 hadronic $W$s should offer excellent opportunities
to probe the considered model. 

We briefly considered a modified $n=1$ model, where the charge two leptons predominantly 
decay into an additional stable, 
neutral state with a mass close to the lightest charge two state. In this case
searches for the charge 2 lepton are more challenging and the current bounds
get relaxed.

In summary, the bounds from direct searches on the new vector leptons in the various scenarios that we have considered, are crucial in constraining the possibility of modifications of the $h \to \gamma\gamma$ and $h\to Z \gamma$ rates. In the case of enhancement of the $h\to\gamma\gamma$ rate, the bounds on the new physics scale from the requirement of the stability of the Higgs potential are slightly less stringent than those obtained in the case of suppression of the $h\to \gamma\gamma$ rate. In general one can achieve 30\% (10\%) modifications of the  $h \to \gamma\gamma$ ($h\to Z \gamma$) rate, due to the effects of vector-like fermions with non-standard hypercharges.  

\section*{Acknowledgments} 

\noindent
We would like to thank Prateek Agrawal, Stefania Gori, Aurelio Juste,
Matthias Neubert, Pedro Schwaller, Carlos Wagner, Lian Tao Wang, and  
Felix Yu for many useful discussions. We also thank David Straub for useful comments.
Furthermore, we would like to thank Rupert Coy for bringing to our attention a mistake in the numerical results of Table I in an earlier version of this paper.
We would like to thank KITP Santa Barbara for warm hospitality and support during completion of this work.
KITP is supported in part by the National Science Foundation under 
Grant No. NSF PHY11-25915.
Fermilab is operated by Fermi Research Alliance, LLC under Contract No.
De-AC02-07CH11359 with the United States Department of Energy. One of us, M.B., is acknowledging the support of 
the Alexander von Humboldt Foundation. The research of W.A. was supported 
in part by Perimeter Institute for Theoretical Physics. Research at Perimeter Institute is supported by the Government of Canada through Industry Canada and by the Province of Ontario through the Ministry of Economic Development \& Innovation.

\appendix
\section{Loop Functions}\label{loopfuncs}

\noindent
The loop functions for the Higgs to diphoton decay read
\begin{alignat}{2}
A_1(x) &= -2-3x-3(2x-x^2)f(1/x) ~~~ & \xrightarrow{x \to \infty} ~~ & -7~, \\
A_{1/2}(x) &= 2x+2(x-x^2)f(1/x) ~~~ & \xrightarrow{x \to \infty} ~~ & 
\frac{4}{3}~, \\
\tilde A_{1/2}(x) &= 2xf(1/x) ~~~ & \xrightarrow{x \to \infty} ~~ & 2~,
\end{alignat}
in which $f(x) = (\arcsin\sqrt{x})^2$ for $x<1$, which is the case relevant for 
us.

\noindent
The functions that enter the approximate expressions for the $h \to Z \gamma$ 
rate read
\begin{alignat}{2}
 h_1(x) &= - \frac{1+2x+2x^2+x^3}{(1-x)^3} - \frac{3x(1+x^2)\log x}{(1-x)^4} 
~~~ & \xrightarrow{x \to 1} ~~ & 0~, \\
 h_2(x) &= - \frac{1+10x+x^2}{3(1-x)^3} - \frac{2x(1+x)\log x}{(1-x)^4} ~~~ & 
\xrightarrow{x \to 1} ~~ & 0~, \\
 h_3(x) &= - \frac{1+11x+11x^2+x^3}{(1-x)^3} - \frac{6x(1+x)^2\log x}{(1-x)^4} 
~~~ & \xrightarrow{x \to 1} ~~ & 0~.
\end{alignat}
 
\noindent
Barr-Zee diagrams lead to the following 2-loop functions in EDMs and MDMs
\begin{alignat}{2}
 g(x) &= \frac{x}{2} \int_0^1 \frac{dy}{y(1-y)-x}
\log\left(\frac{y(1-y)}{x}\right) ~~~ & \xrightarrow{x \to \infty} ~~ & 
\frac{1}{2} \log x, \\
 f(x) &= \frac{x}{2} \int_0^1 \frac{dy(1-2y(1-y))}{y(1-y)-x}
\log\left(\frac{y(1-y)}{x}\right) ~~~ & \xrightarrow{x \to \infty} ~~ & 
\frac{1}{3} \log x~.
\end{alignat}

\section{Couplings}\label{coups}

\noindent
In this appendix we give explicit expressions for couplings of the new lepton 
mass eigenstates with the Higgs as well as with gauge bosons. 
The expressions apply to both the $n=1$ and the $n=2$ case. However, 
in the $n=2$ case, all mixing Yukawas $y_{L\ell}$ vanish and have to be set to 
$0$ in the following expressions.

The mass matrix $\bm{\mathcal{M}}_E$ of the vector-like leptons $E$ can be 
diagonalized by a bi-unitary transformation 
\begin{equation}
 Z_L \bm{\mathcal{M}}_E Z_R^\dagger = \begin{pmatrix} m_1 & 0 \\ 0 &m_2 
\end{pmatrix} ~,
\end{equation}
with $m_1 < m_2$ real and positive.
The most general parametrization of the $Z_L$, $Z_R$ matrices reads
\begin{equation}
 Z_L = \begin{pmatrix} c_L e^{i(\phi_L + \phi_{c_L})} & s_L e^{i(\phi_L +
\phi_{s_L})} \\ - s_L e^{i(\phi_L - \phi_{s_L})} & c_L e^{i(\phi_L -
\phi_{c_L})}\end{pmatrix} ~~,~~~ 
 Z_R = \begin{pmatrix} c_R e^{i(\phi_R + \phi_{c_R})} & s_R e^{i(\phi_R +
\phi_{s_R})} \\ - s_R e^{i(\phi_R - \phi_{s_R})} & c_R e^{i(\phi_R -
\phi_{c_R})}\end{pmatrix} ~,
\end{equation}
with $s_L^2 + c_L^2 = s_R^2 + c_R^2 = 1$. We then denote the left- and
right-handed components of the mass eigenstates by 
\begin{equation}
\chi_L= (P_L \chi_1, P_L \chi_2)^T = Z_L (E_L, \tilde E_L)^T ~,~~ \chi_R
=(P_R \chi_1, P_R \chi_2)^T  = Z_R
(\tilde E_R, E_R)^T ~,
\end{equation}
in which $P_{L,R}=\frac{1}{2}(1\pm\gamma_5)$ are the chiral projection 
operators.

We collect the remaining vector-like lepton $N$ together
with the charged SM leptons into vectors
\begin{align}
 \eta_L=(P_L e, P_L\mu, P_L\tau, P_L N)^T\,,\qquad  \eta_R=(P_R e, P_R\mu,
P_R\tau, P_R N)^T\,,
\end{align}
even though they can only mix in the $n=1$ scenario. 

We parametrize the interactions of $\chi$ and $\eta$ with the Higgs and with 
gauge bosons in the following generic way
\begin{align}
 \Delta \mathcal{L}=&e\,Q_\chi A_\mu \bar \chi \gamma^\mu \chi + e\,Q_\eta A_\mu
\bar
\eta \gamma^\mu \eta\notag\\ 
&+e\,Z_\mu \Big(\bar \chi_L \gamma^\mu\,g_{Z\chi\chi}^L\,\chi_L+\bar\chi_R 
\gamma^\mu
\,g_{Z\chi\chi}^R\,\chi_R ~+~ \bar\eta_L
\gamma^\mu \,g_{Z\eta\eta}^L\,\eta_L+\bar
\eta_R\gamma^\mu
\,g_{Z\eta\eta}^R\,\eta_R\Big)\notag\\
&+\frac{g_2}{\sqrt{2}}\,W_\mu\Big(\bar \chi_L \gamma^\mu
g_{W\chi\eta}^L\,\eta_L+\bar \chi_R \gamma^\mu
g_{W\chi\eta}^R\,\eta_R + \bar \eta_L \gamma^\mu
g_{W\eta\nu}^L\,\nu_L ~+ h.c. \Big)\,\notag\\
&+ h \left(\bar \chi_L \,g_{h\chi\chi}\,\chi_R  \, + 
\bar \eta_L \,g_{h\eta\eta}\,\eta_R ~+h.c.\right)\,.
\end{align}
In the couplings of the photons we have $Q_\eta = -1$ for $\eta = \ell$ and 
$Q_\eta = Q_N$ for $\eta = N$.
For the Higgs couplings with the mass eigenstates $\chi$ we find the following 
expressions
\begin{align}
 g_{h\chi\chi}=\label{ghpsiLpsiR}
 \frac{m_1}{v} &\begin{pmatrix}
s_R^2c_L^2+s_L^2c_R^2&c_Rs_R(c_L^2-s_L^2)\,e^{i(\phi_{c_R}+\phi_{s_R})}\\[4pt]
s_Lc_L(c_R^2-s_R^2)\,e^{-i(\phi_{c_L}+\phi_{s_L})}&
-2
s_Lc_Ls_Rc_R\,e^{-i\phi}
\end{pmatrix}\notag\\[3pt]
+\frac{m_2}{v}&\begin{pmatrix}
-2
s_Lc_Ls_Rc_R\,e^{i\phi}&-\,
s_Lc_L(c_R^2-s_R^2)\,e^{i(\phi_{c_L}+\phi_{s_L})}\\[4pt]
-\,c_Rs_R(c_L^2-s_L^2)\,e^{-i(\phi_{c_R}+\phi_{s_R})}&
s_R^2c_L^2+s_L^2c_R^2
\end{pmatrix}\,,
\end{align}
For the Higgs couplings with $\eta$ we expand in first order in $v^2/M_L^2$ and 
find
\begin{align}
\label{ghetaLetaR}
g_{h\eta\eta}=\begin{pmatrix}
\frac{y_e}{\sqrt{2}}(1-\frac{3}{2}|y_{Le}|^2
\frac{v^2}{M_L^2})&
-\frac{y_e \,y_{Le}^\ast y_{L\mu} v^2}{\sqrt{2}M_L^2}&-\frac{y_e
\,y_{Le}^\ast
y_{L\tau} v^2}{\sqrt{2}M_L^2}&\frac{y_e \,y_{Le}^\ast v}{\sqrt{2}M_L}\\[4pt]
-\frac{y_\mu \,y_{L\mu}^\ast y_{Le} v^2}{\sqrt{2}M_L^2}&\frac{y_\mu}{\sqrt{2}}
(1-\frac{3}{2}|y_{L\mu}|^2\frac{v^2}{M_L^2})&
-\frac{y_\mu \,y_{L\mu}^\ast
y_{L\tau} v^2}{\sqrt{2}M_L^2}&\frac{y_\mu \,y_{L\mu}^\ast v}{\sqrt{2}M_L}\\[4pt]
-\frac{y_\tau \,y_{L\tau}^\ast y_{Le} v^2}{\sqrt{2}M_L^2}&
-\frac{y_\tau \,y_{L\tau}^\ast y_{L\mu} v^2}{\sqrt{2}M_L^2}&
\frac{y_\tau}{\sqrt{2}}(1-\frac{3}{2}|y_{L\tau}|^2
\frac{v^2}{M_L^2})&\frac{y_\tau \,y_{L\tau}^\ast v}{\sqrt{2}M_L}\\[4pt]
\frac{y_{Le}}{\sqrt{2}}
(1-\frac{1}{2}|y_{Le}|^2\frac{v^2}{M_L^2})&
\frac{y_{L\mu}}{\sqrt{2}}
(1-\frac{1}{2}|y_{L\mu}|^2\frac{v^2}{M_L^2})&
\frac{y_{L\tau}}{\sqrt{2}}
(1-\frac{1}{2}|y_{L\tau}|^2\frac{v^2}{M_L^2})&
\sum_{\ell}|y_{L\ell}|^2\frac{v}{\sqrt{2}M_L}
\end{pmatrix}\,.
\end{align}

\noindent
For the couplings of the $Z$ boson with $\chi$ we find,
\begin{align}
 g_{Z\chi\chi}^L&= \frac{s_W}{c_W} Q_\chi \, 1\!\!1 + \frac{1}{2s_W c_W}
 \begin{pmatrix}
 c_L^2 &-c_Ls_Le^{i(\phi_{c_L}+\phi_{s_L})}\\[4pt]
 -c_Ls_Le^{-i(\phi_{c_L}+\phi_{s_L})}&s_L^2 
 \end{pmatrix}\,,\\[3pt]
  g_{Z\chi\chi}^R&= \frac{s_W}{c_W} Q_\chi \, 1\!\!1 + \frac{1}{2s_W c_W}
 \begin{pmatrix}
 c_R^2 &-c_Rs_Re^{i(\phi_{c_R}+\phi_{s_R})}\\[4pt]
 -c_Rs_Re^{-i(\phi_{c_R}+\phi_{s_R})}&s_R^2 
 \end{pmatrix}\,.
\end{align}
The couplings of the $Z$ boson with $\eta$ read
\begin{align}\label{gZetaLetaL}
 g_{Z\eta\eta}^L= \frac{s_W}{c_W} Q_\eta \, 1\!\!1 - \frac{1}{2 s_W c_W} 
\begin{pmatrix}
-1 & 0 & 0
&\frac{2 y_ey_{Le}^* v^2}{M_L^2}\\[4pt]
0 & -1 & 0
&\frac{2 y_\mu y_{L\mu}^* v^2}{M_L^2}\\[4pt]
0 & 0 &
 -1&\frac{2 y_\tau y_{L\tau}^* v^2}{M_L^2}\\[4pt]
\frac{2 y_ey_{Le} v^2}{M_L^2}&\frac{2 y_\mu y_{L\mu} v^2}{M_L^2}&
\frac{2 y_\tau y_{L\tau} v^2}{M_L^2}&1
                   \end{pmatrix}\,,
\end{align}
\begin{align}\label{gZetaRetaR}
 g_{Z\eta\eta}^R= \frac{s_W}{c_W} Q_\eta \, 1\!\!1 - \frac{1}{2 s_W c_W} 
\begin{pmatrix}
 |y_{Le}|^2\frac{v^2}{M_L^2}&\frac{y_{Le}^*y_{L\mu} v^2}{M_L^2}&
\frac{y_{Le}^*y_{L\tau}v^2}{M_L^2}&\frac{v\,y_{Le}^*}{M_L}\\[4pt]
\frac{y_{L\mu}^*y_{Le}v^2}{M_L^2}&|y_{L\mu}|^2\frac{v^2}{M_L^2}&
\frac{y_{L\mu}^*y_{L\tau} v^2}{M_L^2}&\frac{v\,y_{L\mu}^*}{M_L}\\[4pt
]
\frac{y_{L\tau}^*y_{Le}v^2}{M_L^2}&\frac{y_{L\tau}^*y_{L\mu}v^2}{
M_L^2}&
|y_{L\tau}|^2\frac{v^2}{M_L^2}&\frac{v\,y_{L\tau}^*}{M_L}\\[4pt]
\frac{v\,y_{Le}}{M_L}&\frac{v\, y_{L\mu}}{M_L}&
\frac{v\,
y_{L\tau}}{M_L}&1-\sum_{\ell}|y_{N_\ell}|^2\frac{v^2}{
M_L^2}
                   \end{pmatrix}\,.
\end{align}
In the case of the left-handed couplings, in principle also flavor changing 
couplings among the SM leptons are generated at the first order in $v^2 / 
M_L^2$. However, they are additionally suppressed by tiny factors $y_\ell 
y_\ell^\prime$ and therefore completely irrelevant for all practical purposes,
and set to $0$ in~(\ref{gZetaLetaL}).

The couplings of the $W$ boson with $\chi$ and $\eta$ read,
\begin{align}
&g_{W\chi\eta}^L=\begin{pmatrix}
  c_L e^{i(\phi_{c_L}+\phi_L)}&0\\[5pt]
  0&-s_L e^{i(\phi_L-\phi_{s_L})}
 \end{pmatrix}
\begin{pmatrix}
y_e\frac{ y_{Le} v^2}{M_L^2}& 
y_\mu\frac{y_{L\mu} v^2}{M_L^2}&
y_\tau \frac{y_{L\tau} v^2}{M_L^2}&
-1\\[4pt]
y_e \frac{y_{Le}v^2}{M_L^2}&
y_\mu \frac{y_{L\mu} v^2}{M_L^2}&
y_\tau \frac{y_{L\tau} v^2}{M_L^2}&1
                  \end{pmatrix}\,,\\[4pt]
 &g_{W\chi\eta}^R=\label{gWpsiRetaR}\begin{pmatrix}
  c_R e^{i(\phi_{c_R}+\phi_R)}&0\\[5pt]
  0&-s_R e^{i(\phi_R-\phi_{s_R})}
 \end{pmatrix}
 \begin{pmatrix}
\frac{y_{Le} v}{M_L}& 
\frac{y_{L\mu} v}{M_L}&
\frac{y_{L\tau} v}{M_L}&
-1+\frac{1}{2}\sum_\ell\frac{|y_{L\ell}|^2v^2}{M_L^2}\\[4pt]
\frac{y_{Le} v}{M_L}&
\frac{y_{L\mu} v}{M_L}&            
\frac{y_{L\tau} v}{M_L}&
-1+\frac{1}{2}\sum_\ell\frac{|y_{L\ell}|^2v^2}{M_L^2}
                  \end{pmatrix}\,.
\end{align}
Finally also $W$ couplings between $N$ and the SM neutrinos $\nu$ are induced
\begin{equation}
 g_{W N\nu_\ell}^L = - y_\ell \frac{y_{L \ell} v^2}{M_L^2} ~,
\end{equation}
where we neglected neutrino mixing, which is irrelevant for our study.


\end{document}